\documentclass[twocolumn,twocolappendix]{aastex63}

\received{Oct 28, 2021}
\revised{Feb 23, 2022}
\accepted{Feb 25, 2022}
\submitjournal{Icarus}

\newcommand{\Rd}{R_{\rm d}}
\newcommand{\cp}{c_{\rm p}}
\newcommand{\Mb}{M_{\rm B}}
\newcommand{\zb}{z_{\rm B}}
\newcommand{\half}{\frac{1}{2}}
\newcommand{\hstst}{\overline{h}^{**}}
\newcommand{\hsat}{\overline{h}^*}
\newcommand{\senv}{\overline{s}}
\newcommand{\henv}{\overline{h}}
\newcommand{\qsat}{\overline{q}^*}

\newcommand{\CAPE}{{\rm CAPE}}

\newcommand{\qcb}{\overline{q}_{\rm C}}
\newcommand{\qucb}{\overline{q}^{\rm C}}
\newcommand{\quc}{q^{\rm C}}
\newcommand{\qc}{q_{\rm C}}
\newcommand{\qtb}{\overline{q}_{\rm T}}
\newcommand{\qt}{q_{\rm T}}
\newcommand{\qut}{q^{\rm T}}

\newcommand{\Tvb}{\overline{T}_{\rm v}}
\newcommand{\zd}{z_{\rm D}}

\defcitealias{ArakawaSchubert1974}{AS74}
\defcitealias{Moorthi1992}{MS92}
\defcitealias{Sankar2021}{S21}

\graphicspath{{./}{figures/}}

\usepackage{amsmath}
\usepackage{gensymb}
\usepackage{lineno}
\usepackage{nth}

\shorttitle{Modeling convective clouds in Jupiter's $24\degree$ N jet}
\shortauthors{Sankar et al.}

\begin{document}


\title{A new convective parameterization applied to Jupiter: implications for water abundance near the $24\degree$ N region}

\correspondingauthor{Ramanakumar Sankar}
\email{rsankar@umn.edu}

\author[0000-0002-6794-7587]{Ramanakumar Sankar}

\author[0000-0001-6111-224X]{Csaba Palotai}
\affiliation{Florida Institute of Technology \\
150 W University Boulevard \\
Melbourne FL, 32901, USA}

\begin{abstract}
Jupiter's atmosphere features a variety of clouds that are formed from the interplay of chemistry and atmospheric dynamics, from the deep red color of the Great Red Spot to the high altitude white ammonia clouds present in the zones (bright bands in Jupiter's atmosphere). Beneath these upper level clouds, water condensation occurs, and sporadically leads to the formation of towering convective storms, driven by the release of large amounts of latent heat. These storms result in a widespread disruption of the cloud and dynamical structure of the atmosphere at the latitude where they form, making the study of these events paramount in understanding the dynamics at depth, and the role of water in the jovian atmosphere. In this work, we use the Explicit Planetary hybrid-Isentropic Coordinate (EPIC) General Circulation Model (GCM) to study the jovian atmosphere, with a focus on moist convective storm formation from water condensation. We present the addition of a sub-grid scale moist convective module to model convective water cloud formation. We focus on the $24\degree$ N latitude, the location of a high speed jetstream, where convective upwellings have been observed every 4-5 years. We find that the potential of convection, and vertical mass and energy flux of the atmosphere is strongly correlated with the amount of water, and we determine an upper limit of the amount of water in the the region surrounding the jet as twice the solar [O/H] ratio.
\end{abstract}




\section{Introduction and motivation}
\label{sec:intro}
One of the key unknowns in the outer planets is the abundance of heavy elements: carbon, nitrogen, oxygen and sulphur. This information is vital for constraining planet formation models, given the current leading theory that these giant planets form from core accretion followed by the attraction of a gaseous envelope \citep{Pollack1996,Atreya2019}. However, in order to grow and maintain the current heavy element abundance, these planets must have formed closer to the Sun and migrated outwards. The difference between the specifics of these planet formation models depends strongly on the resulting metallicity (heavy element to hydrogen ratio) of the planet, and thus, constraining these models requires precise values of heavy element abundances.
The vast differences in metallicity between the four outer planets is an unsolved mystery: on Jupiter, the carbon abundance is thought to be approximately equal to the solar enrichment value (i.e., $1\times$ solar [C/H] ratio), while more than $40\times$ solar on Uranus and Neptune. Therefore, atmospheric carbon forms a thin stratospheric haze of hydrocarbons on Jupiter and Saturn \citep{Friedson2002}, but opaque, white clouds on the ice giants \citep{WeidenschillingLewis1973}. The effect of elemental abundance on meteorology is a valuable relation that we can use to infer their values, given the difficulty in obtaining these measurements from observational data. 

This is especially true on Jupiter, where the clouds form as a result of several different species being affected by chemistry, photochemistry, condensation and local atmospheric dynamics. Among these are three major cloud types, forming from ammonia (NH$_3$), ammonium hydrosulfide (NH$_4$SH) and water (H$_2$O) \citep{WeidenschillingLewis1973,AtreyaWong1999}. The former two are responsible for visible cloud formation, while the latter forms in the deeper atmosphere. Remote sensing provides data on the upper layer, but observing the deeper atmosphere is difficult due to obstruction by the upper level clouds. As such, we infer the properties of water and fluid dynamical processes in the deep atmosphere from their effect on upper level clouds. Furthermore, the amount of water on Jupiter is a long-sought value, with several studies having determined only localized values from spectroscopic features \citep[e.g.,][]{Bjoraker2018,Li2020}. Deep water abundance is strongly tied to the intensity and frequency of convective outbreaks \citep{Guillot1995,Li2015,Leconte2017}, as these convective storms are fueled by the latent heat release of water, much like the tropical storms on Earth. 

\citet{Gehrels1974} used {\it Pioneer} data to determine the ratio of emitted to absorbed radiative flux from Jupiter, and interpreted the results to show a convectively dominated interior. A few years later, images from the {\it Voyager} spacecrafts showed bursts of cloud formation that occur periodically in time \citep{Smith1979,Hunt1982}. This convective activity was attributed  to a deep atmospheric wave, where the crests cause periodic upwelling (plumes), forcing volatiles from within the atmosphere to above the cloud tops. The {\it Galileo} spacecraft observed moist convective activity along with lightning at pressures of $p \gtrsim 3$ bar \citep{Gierasch2000}, where only water is expected to condense. These storms led to the formation of towering clouds over $50$ km high, with a mechanism similar to the development of mesoscale convective complexes on Earth. 
These events are thought to be driven by the release of Jupiter's internal heat, which is remnant from the gravitational collapse during planet formation. The convective storms are responsible for carrying a significant fraction of this heat flux to the upper atmosphere \citep{Gierasch2000}. 
\citet{Ingersoll2000} argue that this energy helps sustain the zonal (i.e., east-west) wind structure, making these disturbances critical in the study of jovian atmospheric dynamics. 

The ascent of a moist parcel of air is driven by the release of latent heat, i.e., condensation of cloud particles heat up the parcel and make it more buoyant compared to the surrounding air. \citet{Stoker1986} argued that, given a solar composition of elements on Jupiter, ammonia and ammonium hydrosulfide simply do not have the potential (both in terms of latent heat release, and density of clouds) to provide the energy needed for such large convective plumes. Therefore, the likely source of these plumes is from the deep water cloud. Using observations of plume formation in the South Equatorial Belt (SEB), \citet{Fletcher2017} found that the thermal emission centered around the $5~\mu$m wavelength before and during the upheaval of volatiles is consistent with a deep cloud base, near that of the expected water cloud. 

Furthermore, the intensity and frequency of convective storms on gas giants is strongly tied to the amount of water. Ideally, the strength of convection should scale linearly with the water abundance, since increasing the water content increases the latent heat release from condensation, leading to higher convective potential \citep{Stoker1986}. However,  on Jupiter, water is heavier than the dry air (the molar mass ratio is $\sim 8$), and thus increasing the water content in the deeper atmosphere increases the stability of the upper atmosphere due to this mass-loading effect. Therefore, there is a drop-off in moist convective activity as the mass loading effect increases, and theoretical calculations show that water-based moist convection should cease for a H$_2$O deep abundance greater than about 5-10$\times$ solar \citep{Guillot1995,Li2015,Leconte2017}.
This work focuses on this aspect of the jovian atmosphere: the link between the deep water abundance and convective storm formation, in an effort to better constrain the mass of water on Jupiter and characterize the processes that lead to storm outbreaks. We use the Explicit Planetary hybrid-Isentropic Coordinate (EPIC) General Circulation Model (GCM) \citep{Dowling1998,Dowling2006} to study the formation of convective storms on Jupiter, with a focus on constraining the water content, and the relation between the small scale convective events and the large scale cloud morphology and atmospheric dynamics. The goal of this study is to investigate the effect of the deep abundance of water and ammonia on the convective capability and dynamic structure of the atmosphere, in an effort to constrain the composition of the jovian atmosphere.

\begin{figure}
	\includegraphics[width=\columnwidth]{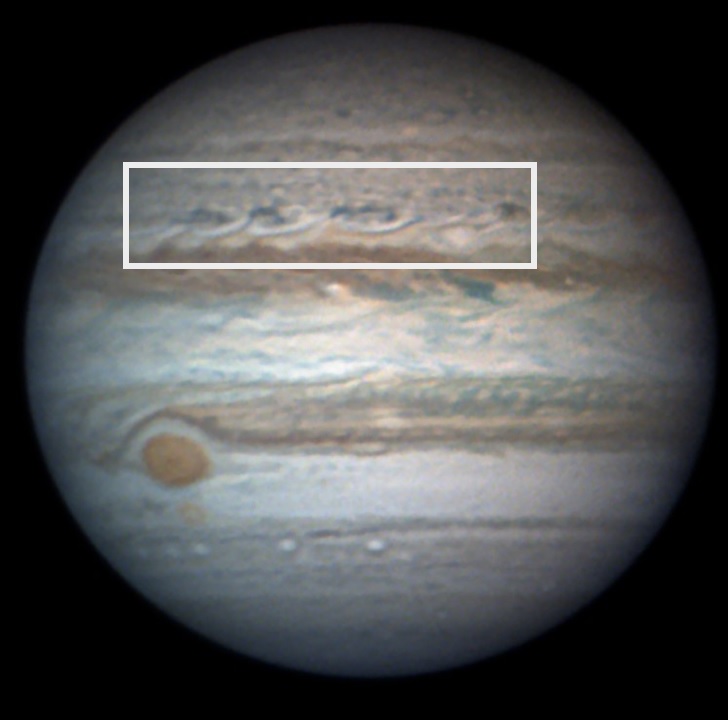}
	\caption{Observations of alternating bright and dark cloud pattern in the jet after a convective outbreak in 2016 \citep{SanchezLavega2017}. Credit: L~Dauvergne/Fran\c{c}ois~Colas/S2P/IMCCE/OMP}
    \label{fig:wave_obs}
\end{figure}

Previous studies have used moist convective models to analyze the development of plumes. For investigating the role of the atmospheric structure on moist convective intensity, \citet{DelGenio1990} use a moist convective parameterization based on the  \citet{ArakawaSchubert1974} scheme. They study the effect of varying water abundances and deep temperature profile on the resulting moist convective intensity, and the final quasi-equilibrium steady state using a 1-dimensional model.  They find that static stability in the environment is mostly  affected by latent heat release rather than molecular weight effects, and the environment quickly reaches a steady state that is neutrally stable to moist convection. However, they do note that more sophisticated cloud microphysics will need to be implemented to fully study the effect of molecular weight effects on static stability. 

\citet{Hueso2001} used a three dimensional non-hydrostatic model to create a localized region ($\sim200$ km$\times200$ km) which was perturbed to initiate plume development. They consider a simple diagnostic cloud scheme where supersaturated vapor is immediately converted to ice / liquid, which is then subject to sedimentation by a factor $f_c$. As expected, they found that initial water vapor content (defined by both the deep abundance and the initial relative humidity) was a strong factor in the intensity and the vertical extent of the storm. The small size of their grid made it impossible to study the formation of the larger bursts observed, but the energy released from convective instabilities formed by a cluster of such storms could explain the observed heat flux.

More recently, \citet{Sugiyama2014} ran a large-scale ($\sim 1000$ km horizontally) 2-dimensional simulation with an explicit treatment of convection and condensation (of water, ammonia and ammonium hydrosulfide). They force the atmosphere with upper level cooling to trigger dynamical instability and find that storm formation is periodic in time, with a marked difference in the atmospheric structure during the outbreaks and the interim quiescent periods. The time period between the outbreaks increases with water abundance.

In this work, we focus on the 24$\degree$ N jet region to study the effects of the convective outbursts and to understand the observed cloud structure following these convective outbreak events. 
Observations of jovian cloud morphology 
have shown that, at this latitude, near the fastest eastward jet on Jupiter, convective outbreaks occur every 4-5 years \citep{Fletcher2017b, Rogers2019}. Each upwelling leads to a widespread disturbance in the chromophoric, cloud and wind structure of the jet \citep{ SanchezLavega1991, SanchezLavega2008,SanchezLavega2017,Rogers2020,PerezHoyos2020}. Figure~\ref{fig:wave_obs} shows the cloud features associated with such outbreaks. These wave-like cloud features have a speed less than that of the jet, and are composed of an alternating pattern of bright clouds and dark, cloud-free regions \citep{SanchezLavega2017}. This wave is therefore strongly correlated with the cloud formation within the jet, and is key in understanding the cloud structure and convective outbreaks in this region, given that they form in the immediate aftermath of convective plumes.

Consequently, studying these storms is vital in understanding the structure of the deep atmosphere of Jupiter and their effects on the upper level clouds. 
To do so, we use the EPIC model, and update it with a moist convective scheme based on \citet[][hereafter MS92]{Moorthi1992}, so that we can physically model the effects of moist convection in the hydrostatic framework. Prior work on this region has shown that convection and cloud formation within the jet after the initial outbreak is driven by an upper level (above 190 hPa) potential vorticity (PV) wave, which forms after the initial plume perturbs the atmosphere \citep[][hereafter S21]{Sankar2021}. In this paper, we augment this previous model with the addition of the moist convective scheme in order to directly infer the convective ability and effects of convection in this region.

In Section~\ref{sec:mcscheme}, we detail the addition of the moist convective scheme and the changes made from the original \citetalias{Moorthi1992} version. In Section~\ref{sec:testcases}, we describe our
test cases that we used in order to validate the scheme, and to obtain a holistic view on how convection 
works in the jovian atmosphere. Finally, we apply our new scheme to 
the $24\degree$ N latitude region to simulate the aftermath of the convective outbreak, in Section~\ref{sec:3dsim}.

\section{Model description and test strategies}

\subsection{EPIC model and stratiform cloud scheme}
EPIC is an atmospheric model that solves the primitive equations on an oblate spheroid \citep{Dowling1998,Dowling2006} on a hybrid vertical coordinate $\zeta$ \citep{KonorArakawa1997}, which transitions from the potential temperature ($\theta$) near the top of the model to scaled pressure ($\sigma$) near the bottom. 

EPIC uses a bulk cloud microphysics parameterization \citep{Palotai2008} based on \citet{HDC2004} for stratiform clouds.  Five phases are studied for each species: vapor, solid (ice cloud), liquid (cloud droplets), snow and rain. 
Ice particles are assumed to be porous and bullet-shaped, with a density half that of bulk ice, while snow is treated as hexagonal plates.
The cloud scheme explicitly calculates the transition of particles between these phases. These processes are split into those that are taken to be instantaneous (melting, freezing), and non-instantaneous (condensation, initiation, deposition, autoconversion, collection and evaporation). 

Sedimentation velocity is calculated beforehand for different particle sizes based on \citet{PruppacherKlettbookCh10}, and fit to a power law as a function of diameter for different pressures, ensuring an accurate and quick treatment of precipitation \citep{Hadland2020}. On Jupiter, these are derived for water and ammonia ice, rain and snow. 
The fits used here are the same as in \citetalias{Sankar2021}.

\subsection{Relaxed Arakawa-Schubert scheme}
\label{sec:mcscheme}
Studies of Earth's atmosphere has made great strides in studying convection, 
even within the hydrostatic framework that is used by most numerical atmospheric
models. Cloud processes on Earth are fueled by insolation, and by the dynamics near
the turbulent boundary layer. Thus the representation of moist convective events in models is very important to model the local/global water and energy budgets. \citet{Arakawa2011} and 
\citet{Arakawa2016} present an overview and history of moist convective treatment
in different types of atmospheric models. Convective schemes fall under two large umbrellas: the convective adjustment scheme is one where the effects of convection are diagnosed and the modifications are then made to the large (grid) scale variables to compensate. Conversely, cloud resolving models (CRMs) use high resolution grids to explicitly resolve convective  clouds. 
For planetary applications, both types of models exist -- EPIC falls under the former category due to its hydrostatic nature and large ($\gtrsim 100$ km)
grid size, while models developed by \citet{Hueso2001} represent the localized
non-hydrostatic approach of CRMs. However, EPIC previously lacked a treatment 
of convective clouds, while the latter does not have an explicit treatment of cloud microphysics. Furthermore, given the expansive surface area on the gas giants,
grid sizes in atmospheric models of these planets are usually at least an order of magnitude
larger than they are on Earth. Therefore, CRMs on gas giants will undoubtedly 
be unable to cover large areas and remain computationally efficient, 
and thus, are unsuitable for studying planetary scale cloud formation. 

Consequently, we augment the EPIC GCM with a sub-grid scale convective adjustment scheme. This will allow the model runs to remain computationally viable for studies of planetary scale cloud formation, while also providing a framework to study convective clouds. In this study, we implement the Arakawa-Schubert style of convective parameterization. However, in its base state \citep[][hereafter AS74]{ArakawaSchubert1974}, the closure assumption is generally too strict, even for Earth meteorology \citep{Yano2020}. This is due to the lack of explicit coupling between cloud formation and large scale dynamics. Therefore, we instead use the Relaxed Arakawa-Schubert \citepalias[RAS,][]{Moorthi1992} which relaxes the sounding towards the equilibrium after each timestep, rather than explicitly solving for the quasi-equilibrium state. In the following subsections, we will detail the formalism of this scheme, with particular note on our adaptation to the EPIC model. Note that we will be implementing the RAS v2 \citep{Moorthi1999} rather the version detailed in the original paper, although the general formalism is the same.

\subsubsection{Cloud ensemble, cloud types and closure scheme}
The RAS uses the same fundamental framework as all Arakawa-Schubert (AS) schemes. The general
principle is to calculate all possible clouds within the vertical grid column,
and determine the updraft profile, and subsequent changes to the environmental variables as a result
of the updraft. All sub-grid scale clouds are assumed to have a base at the same level $\zb$, but have varying cloud top altitudes $\zd$ (the detrainment level), which, in the model, can start anywhere above the first layer up from $\zb$. Therefore, several clouds (or `cloud types') can co-exist within the same grid column in our model, and the $i$th cloud is defined as one where the $\zd$ for cloud $i$ lies on the $i$th vertical model layer. 

Convection is driven by the idea of buoyancy: pockets of air with lower density than the surrounding will freely rise, whereas those that are denser will sink. Therefore, convective cloud updrafts are driven by having positive buoyancy, and thus, a cloud that has a top at $\zd$ is positively buoyant between $\zb$ and $\zd$ and has neutral buoyancy at $\zd$ (the cloud must stop convecting freely when the positive buoyant forcing disappears). For each cloud invoked, the RAS scheme determines whether this cloud will be convective based on this buoyancy condition, and determines the updraft profile such that the buoyancy profile within the updraft matches this constraint. 
A second criteria for convection involves the conservation of energy within the grid cell. Convective upwellings must use the energy that is available for convection within the grid cell and thus, the ensemble of clouds within the grid cell must be limited by the total convective potential energy of all possible clouds within the column. 


\subsubsection{Vertical profiles of mass and energy flux, and buoyancy}
Following \citetalias{Moorthi1992}, we calculate the buoyancy of a parcel within the updraft as,

\begin{equation}
    \label{eq:buoyancy_main}
    B = g\left[\dfrac{T_{\rm v} - \overline{T}_{\rm v}}{\overline{T}_{\rm v}} - \left(\qc - \qcb\right)\right],
\end{equation}
where $g$ is the local acceleration due to gravity, $T_{\rm v}$ is the virtual temperature, $\qc$ is the cloud specific humidity (ratio of cloud mass density to total density). Overbars denote atmospheric (grid scale) values, while variables without the overbars are in-cloud (sub-grid scale) values. The first term corresponds to the traditional buoyancy term (i.e. difference in density), while the second terms adds the contribution (i.e. reduction in buoyancy) from the weight of the in-cloud condensates. 

Within the cloud, upwelling carries energy and mass to the upper levels, and thus the in-cloud temperature and cloud mixing ratios are determined from the vertical updraft profile. 
Generally, the vertical mass flux profile can be a complex function of the environmental variables but for simplicity, we follow \citetalias{ArakawaSchubert1974} and assume that the updraft entrains a constant fraction of mass flux $\lambda$, within a layer $\delta z$, 

\begin{equation}
	\dfrac{dM}{dz} = \lambda M(z).
\end{equation}
We can normalize the mass flux by the value at the cloud base, such that $M(z) = \Mb \eta(z)$, 
where $\Mb$ is the mass flux at the cloud base and $\eta(z)$ is the normalized mass flux profile. Therefore, the mass flux relation above simplifies to

\begin{equation}
	\dfrac{d\eta}{dz} = \lambda \eta(z),
\end{equation}
In this fashion, $\lambda$ and $\Mb$ completely define a given updraft, where $\lambda$ defines the fraction of vertical mass flux that is entrained from the surrounding environment at any given level, and $\Mb$ defining the strength of the updraft. Therefore, we can solve for the normalized profile by integrating the differential equation. We integrate from the cloud base ($\zb$, where the updraft begins), up to a level $z$, that is above the base,

\begin{equation}
	\label{eq:eta_equation}
	\dfrac{1}{\eta(\lambda)}\dfrac{d\eta(\lambda)}{dz} = \lambda  \Rightarrow 
	\eta(\lambda, z) = \exp\left(\lambda (z - \zb)\right),
\end{equation}
where $\zb$ is the altitude of the cloud base. This ensures that $\eta(\lambda, \zb) = 1$, 
such that $M(\lambda, \zb) = \Mb$. However, in this form, solving for $\lambda$ is difficult. 
The RAS scheme instead approximates $\eta(\lambda, z)$ as,

\begin{equation}
	\label{eq:eta_RAS}
	\eta(\lambda, z) = 1 + \lambda \zeta + \lambda^2 \xi,
\end{equation}
where $\zeta = z-\zb$ and $\xi = \frac{1}{2}\zeta^2$. This quadratic approximation reduces the computational cost of determining the updraft mass flux profile, while retaining the necessary accuracy. 

With the given updraft parameterization, we solve for the vertical profiles of mass and energy with the vertical continuity equation. We define the dry ($s$), moist ($h$) and moist saturated ($h^*$) static energies, as,

\begin{eqnarray}
	s & = & \cp T + \Phi, \label{eq:s}\\
	h & = & s + L q, \label{eq:h}\\
	h^* & = & s + L q^* \label{eq:hstar},
\end{eqnarray}
where $\cp$ is the specific heat capacity at constant pressure, $T$ is the gas temperature, $\Phi$ is the geopotential, $L$ is the latent heat per unit mass, and $q$ is the vapor 
specific humidity of the moisture species in the air. $h^*$ is the saturated moist static energy, which is similar to $h$, but defined from $q^*$, which is the saturation vapor specific humidity at a given pressure and temperature, i.e., the specific humidity at which the atmosphere is fully saturated. During adiabatic motion, $s$ is a conserved quantity, 
while during moist ascent, $h$ is conserved. 

\begin{figure*}[ht]
	\includegraphics[width=\textwidth]{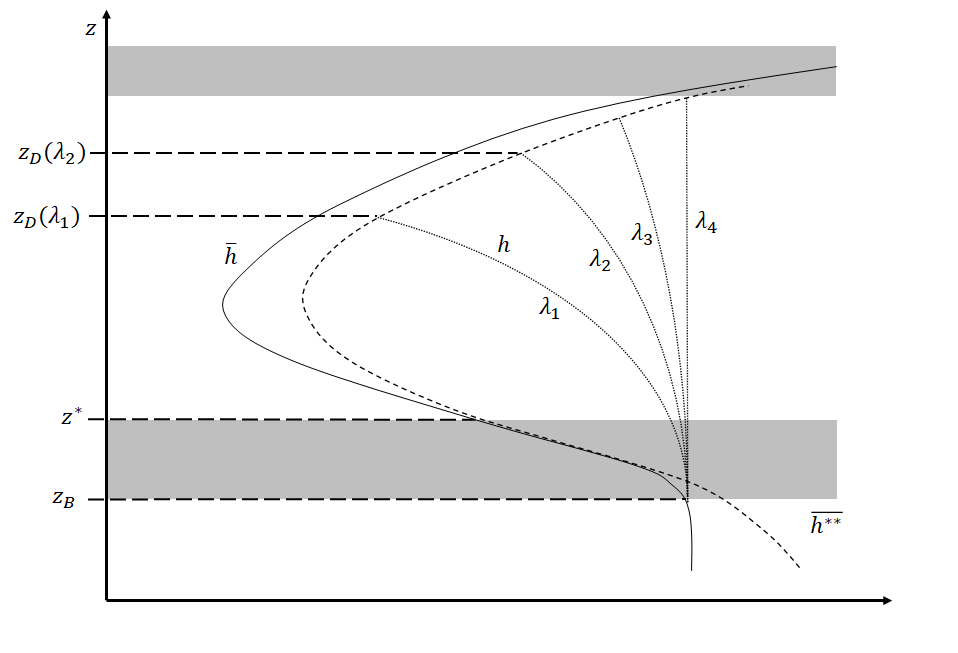}
	\caption{A idealized view of the updraft profile and boundary condition. The solid line shows the environment moist static energy ($\overline{h}$), while the dashed line shows the grid scale virtual moist static energy ($\hstst$). The dotted lines show different updraft moist static energy profiles ($h$), with decreasing entrainment, $\lambda_1 > \lambda_2 > \lambda_3 > \lambda_4$. The updraft starts at $\zb$ and has a top at each $\zd$. At this top the updraft $h$ must match $\hstst(\zd)$. The grey areas show regions where such an updraft is impossible since the updraft can never entrain lower moist static energy than at the detrainment level for the lower gray region. For the lower region, $z^*$ represents the altitude below which the updraft would require an unphysically large entrainment to achieve neutral buoyancy (we currently limit the entrainment parameter to $10^{-4}$ m$^{-1}$ as detailed in the text below). For the upper region, the updraft can never entrain moist static energy that is higher than what it started with at the cloud base. Here, we are ignoring the small mass loading effect for simplicity.
	}
	\label{fig:updraft_h_profile}
\end{figure*}

With these definitions, we can write the cloud and atmospheric virtual temperatures $T_{\rm v}$ and $\Tvb$ in terms of the moist static energy. Therefore, the first condition (neutral buoyancy at the cloud top), then becomes, using the definition of the buoyancy from Equation~\ref{eq:buoyancy_main},

\begin{equation}
    \label{eq:cloud_top_buoyancy}
    B(\zd) = h(\zd) - \hstst(\zd) - \tilde{L}\left(\qc(\zd) - \qcb(\zd) \right) = 0,
\end{equation}
where $\hstst$ is the grid-scale virtual moist saturated static energy, and $\tilde{L}$ is latent heat term in Equation~\ref{eq:h}, accounting for virtual effects. 
Effectively, in the limit of low cloud condensate density, the cloud top buoyancy condition is matched when the updraft moist static energy profile ($h$) matches the grid-scale virtual moist saturated static energy profile ($\hstst$). Figure~\ref{fig:updraft_h_profile} shows a schematic of how this cloud top buoyancy condition is matched at different heights. Since $\henv$ is usually smaller than $\hstst$ for a subsaturated atmosphere, increasing the entrainment from the surrounding generally decreases $h$ within the updraft, shifting the curve to the left. The value of $\lambda$ is the slope of $h$ that intersects $\hstst$ at $\zd$. 

With this parameterization of $\eta$, we can determine the cloud top buoyancy condition ($B(\zd)=0$), which gives, 

\begin{equation}
    a \lambda^2 + b \lambda + c = 0,
\end{equation}
where $a$, $b$ and $c$ are functions of only the grid-scale variables (i.e. model values). This gives two solutions for the entrainment parameter $\lambda$, where we choose the highest positive value that is less than $10^{-4}$ m$^{-1}$ (i.e. an entrainment of 10\% mass flux per kilometer of updraft). This is an arbitrary value currently chosen to maintain numerical stability in our model. This upper limit of $\lambda$ essentially prohibits very shallow updrafts, as shown by the lower grey region in Figure~\ref{fig:updraft_h_profile}. This parameter can be tweaked to optimize the false positive rate of updraft triggers for shallow convection. Generally, for an atmospheric profile similar to the one shown in Figure~\ref{fig:updraft_h_profile}, where $\henv$ decreases with height until near the tropopause, it is impossible to get two positive solutions, since $h$ will generally decrease monotonically with height, making both $a$ and $b$ negative. The derivation of $a$, $b$ and $c$ are given in \citet{Moorthi1999}, and in the Appendix, along with the vertical tendencies for the moisture and thermodynamic variables. 

\subsubsection{Energy conservation}
Prior to calculating the updraft tendencies, the cloud base mass flux, $\Mb$ must be determined. We need to determine $\Mb$ which describes the strength of the convective upwelling, through a closure (given by energy conservation) for our system of equations. Convection is triggered by the increase in convective available potential energy (CAPE) or the decrease in convective inhibition (CIN), through external forcing (i.e., large-scale dynamics). CAPE defines the total potential energy gained by a parcel when it rises through a region of positive buoyancy,

\begin{equation}
    \label{eq:CAPE}
    {\rm CAPE} = \int_{\zb}^{z_{\rm top}} B(z) dz,
\end{equation}
while CIN defines the total energy required by the parcel to rise through a region of negative buoyancy to get the base of the upwelling layer,

\begin{equation}
    {\rm CIN} = -\int_{z_0}^{\zb} B(z) dz.
\end{equation}

Therefore, the trigger function for convection within a column is tied to change in buoyancy for clouds of different heights. We implement this using the idea of large-scale changes in atmospheric CAPE, similar to \citet{Zhang2002}. The change in CAPE for cloud type $i$ due to cumulus effects is obtained by integrating the change in buoyancy with height,

\begin{equation}
	\left(\dfrac{d\CAPE_i}{dt}\right)_{\rm cu} = \int_{\zb}^{\zd} \dfrac{dB_i}{dt} dz
	= -M_{b,i} F,
\end{equation}
where $F$ defines the fractional change in CAPE for cloud type $i$ due to a unit mass flux, $F$, and the cloud top height ($z_{\rm top}$) in Equation~\ref{eq:CAPE} is replaced with the detrainment level for cloud type $i$. We determine the large scale effects on the CAPE for cloud type $i$ using finite difference,

\begin{equation}
	\label{eq:dCAPEdt_LS}
	\left(\dfrac{d\CAPE_i}{dt}\right)_{\rm LS} \approx \dfrac{\CAPE_i(t+\Delta t) - \CAPE_i(t)}{\Delta t}.
\end{equation}
Therefore, we can then determine the cloud base mass flux, as,

\begin{equation}
	\Mb = -\dfrac{1}{F} \left(\frac{d\CAPE_i}{dt}\right)_{\rm LS}.
\end{equation}
The derivation of the CAPE formalism used  in this scheme, and the equation for $F$ are given in the Appendix.

\subsubsection{Relaxation to quasi-equilibrium}
The distinction between the AS and RAS scheme is that the RAS relaxes the sounding towards a quasi-equilibrium over several timesteps, rather than within a single one. This is due to the fact that solving directly for equilibrium is generally too strict for convective parameterization \citep{Yano2020}. Consequently, we only allow a fraction $\alpha$ of the mass flux to affect 
the sounding. This acts similarly to, for example, under-relaxed Jacobi's iteration method for linear system solvers, where
the true solution to the system of equation is obtained after several iterations. 

In the case of moist convection, this means that the quasi-equilibrium will be achieved after $n$ timesteps, 
or over a timescale of $\tau=n \Delta t$, where each timestep steps the model forward by $\Delta t$. $\tau$ is then the timescale over which inter-cloud interactions 
self-stabilize the large-scale forcing. Therefore, with this principle, we can define the relaxation parameter, 
$\alpha$, as,

\begin{equation}
	\alpha = \dfrac{1}{n} = \dfrac{\Delta t}{\tau},
\end{equation}
where $\tau \sim 10^3 - 10^4$ s on Earth \citepalias{ArakawaSchubert1974}. On Jupiter, this is currently unknown
from observations. However, modeling attempts show that these convective storms stabilize on the 
timescale of ${\sim} 10$ hours \citep{Hueso2001}.
Therefore, we test different values of $\tau$  to see
the sensitivity of the parameter to jovian convection. On Earth, increasing $\tau$ linearly increased the amount of iterations required to relax the model to the critical value of CAPE
\citepalias{Moorthi1992}, so we expect the process to work similarly on Jupiter. 

\begin{figure}[h]
	\includegraphics[width=\columnwidth]{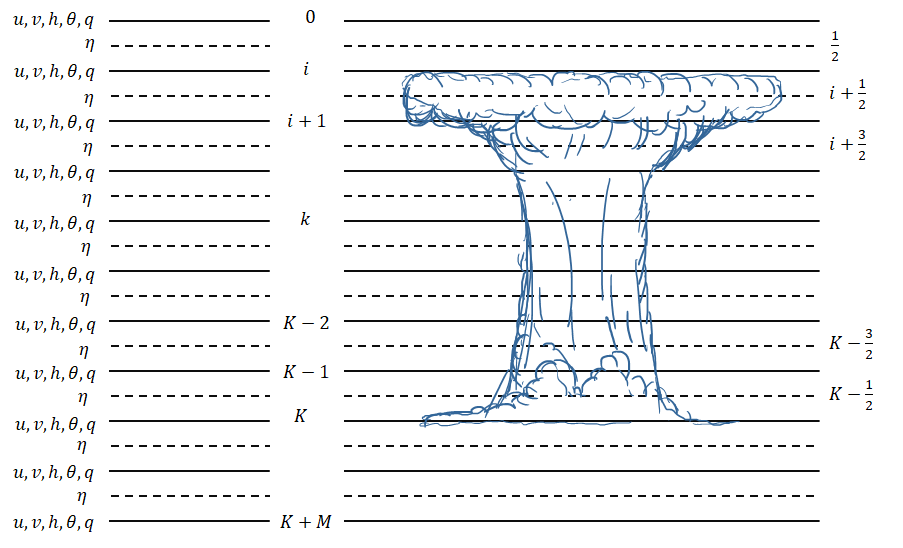}
	\caption{Vertical discretization of a single vertical column (i.e., a horizontal grid cell) in the model, denoting the indices that correspond to different levels of the updraft. Integer indices correspond to cell edges, as is the convention in EPIC since thermodynamic variables are stored	on the edges. The cloud base is located at $K$ and detrains at $i$. Half-integer values correspond to fluxes. The top of the atmosphere corresponds to index $0$ and the bottom is $M$ layers below $K$, where $M\geq 1$. 
	}
	\label{fig:vertical_discrete}
\end{figure}

\subsubsection{Vertical discretization}
The model is discretized vertically as shown in Figure~\ref{fig:vertical_discrete}. EPIC features a vertically staggered grid, with thermodynamic and dynamic variables stored on the grid interface (i.e. top/bottom of the cell) rather than the center, and so that index $k$ denotes the $k$th edge from the top (Figure~\ref{fig:vertical_discrete}). $K$ defines the cell edge corresponding to the bottom of the cloud, i.e., $z_{K} = \zb$, and $i$ denotes the grid interface corresponding to the cloud top. In the original RAS scheme, values are stored in the cell center, and the cloud starts at the cell center $K$ and detrains at the center $i$. However, in EPIC, we retain the numbering scheme but note that integer values correspond to edges, rather than centers. Correspondingly, the mass fluxes are defined on the cell centers in EPIC, (i.e., at half-integer indices), while the tendencies (mass and heat) are located on the cell edges (integer indices) so as to be able to update the thermodynamic variables. $\eta_{K-\half} = 1$ since that is the base of the updraft and $\eta_{i-\half} = 0$ because the updraft stops at $i$. Thermodynamic values at half-integer indices are interpolated similarly to \citet{KonorArakawa1997}. The vertical discretization is shown in Figure~\ref{fig:vertical_discrete}.

\subsection{1-dimensional test cases}
\label{sec:testcases}
\begin{figure}[ht]
	\includegraphics[width=\columnwidth]{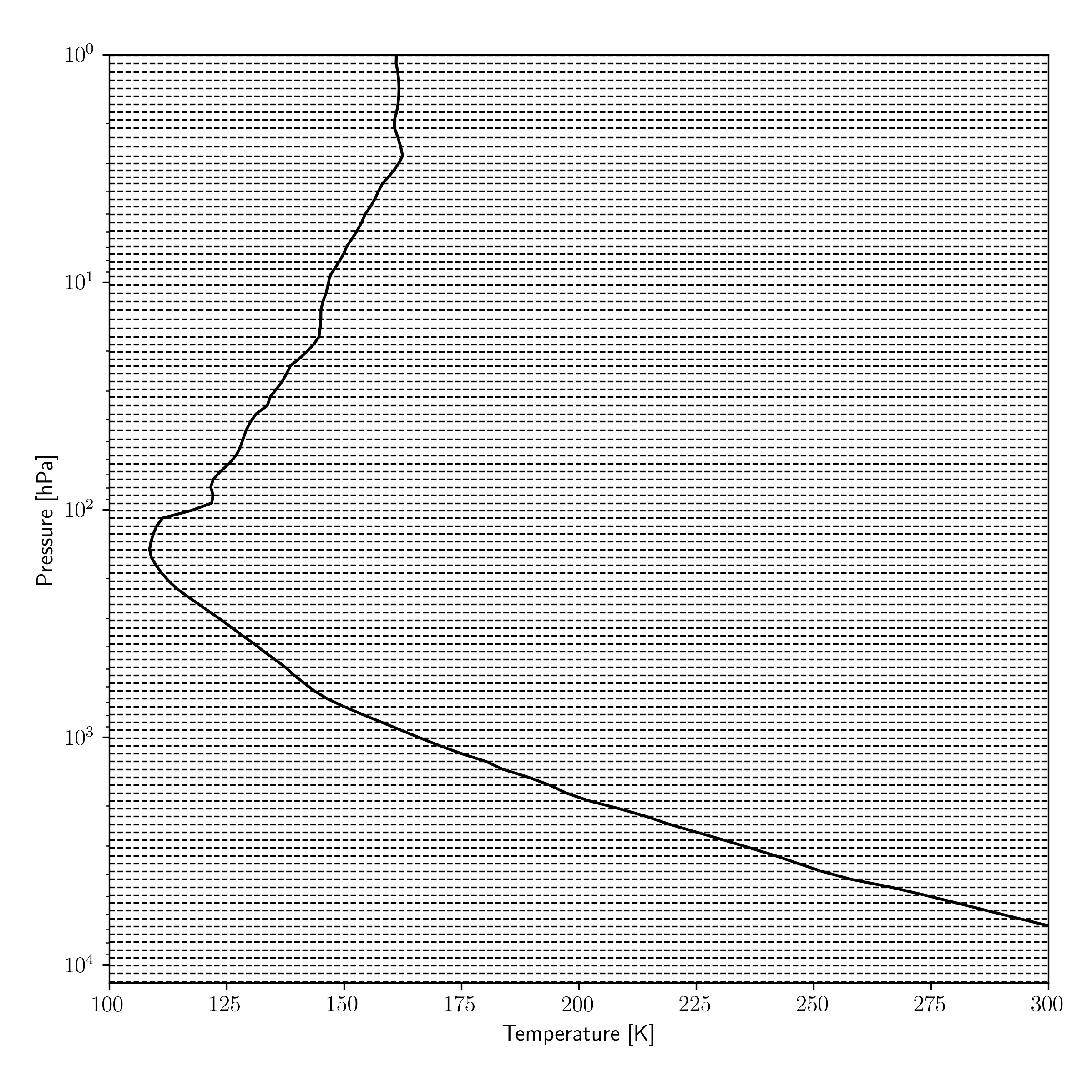}
	\caption{Temperature-pressure profile of the jovian used in the 1-D test cases from \citet{Moses2005}. The dashed lines show the locations of the vertical layers.}
	\label{fig:1d_temp_profile}
\end{figure}

To test the moist convection scheme, we will run one-dimensional (vertical only) simulations of the jovian atmosphere. The objective of these test cases is to investigate how the new module functions on gas giant atmospheres, in a simplified, controlled setting. 
With this model, we study the profiles of the variables that affect moist convection and ensure that the new code follows the energy and mass conservation principles defined in Section~\ref{sec:mcscheme}. We also test to make sure that cloud top buoyancy condition is met properly. In particular, we are interested in the updraft profile of the moisture and energy variables, as well as a test of the cloud top buoyancy condition. In the following subsections, we look at how these updraft profiles change during a convective event, and their effect on the grid-scale variables. The initial vertical temperature-pressure profile for the jovian atmosphere, that we use in our model, is shown in Figure~\ref{fig:1d_temp_profile}. 

\begin{figure*}[ht]
	\includegraphics[width=\textwidth]{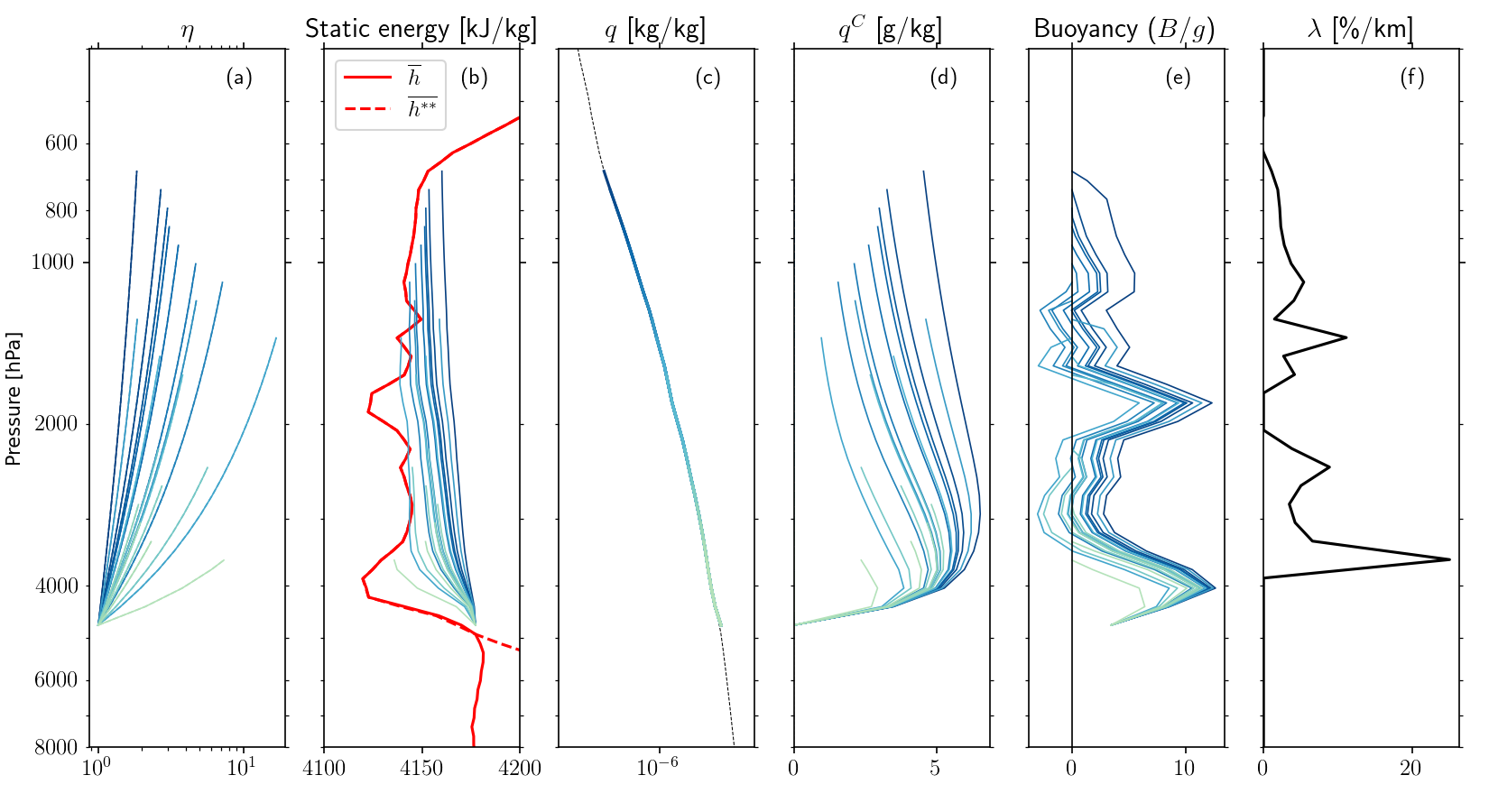}
	\caption{Output profile from the first calculation of the moist convective updrafts in EPIC. In each panel, the thin blue-green colored lines correspond to a cloud type. Panel (a) shows the normalized mass flux profile for each cloud type. Panel (b) depicts the moist static energy $\henv$ in solid red and virtual moist static energy $\hstst$ in dashed red. The color lines show the moist static energy with the updraft ($h$). Panel (c) corresponds to the vapor profile within updraft, with the thin dashed line showing the environmental saturation vapor pressure. Panel (d) shows the cloud condensate profile within the updraft, with solid corresponding to ice and dashed corresponding to liquid. At these temperatures, there are no liquid cloud particles within the updraft. Panel (e) shows the buoyancy of the updraft normalized by $g$, and panel (f) denotes the corresponding entrainment parameter. }
	\label{fig:1d_updraft_profile}
\end{figure*}

Figure~\ref{fig:1d_updraft_profile} shows the vertical profiles of the updraft variables
after the first timestep in the model, with each thin colored line referring to a different cloud type. 
Panel (a) shows the normalized mass flux profile for different cloud types (i.e., clouds within the vertical column that detrain at different levels). Each cloud type (i.e., colored line in Figure~\ref{fig:1d_updraft_profile} is a different convecting cloud within this atmospheric column). Generally, the shallower
updrafts require more entrainment of dry air (i.e., larger entrainment parameter $\lambda$, producing correspondingly larger mass flux $\eta$) so as to satisfy the neutral buoyancy condition at the cloud top, since larger entrainment of dry air shifts the $h$ towards the $\henv$ profile more quickly. This is consistent with the vertical moist static energy profile shown in panel (b). The cloud base is at the layer where grid scale moist static energy decreases (as vapor condenses, $\henv$ decreases correspondingly). Therefore, to reach the cloud top closest to the base, the cloud has to entrain much more dry air (with lower $\henv$) compared to an updraft that reaches the $1$ bar level. The air within the updraft is saturated and closely follows the saturation vapor pressure curve (thin dashed line in the panel c). 

More importantly, the buoyancy profile (panel e) matches the cloud top condition. At the top of each 
updraft, the buoyancy reaches a value of zero to machine precision, about $10^{-16}$. Note, however, that due to the cloud mass
this does not exactly correspond to $h(z_D) = \hstst(z_D)$, as the curves do not intersect in the panel (b).
Instead, the small correction is due to the mass loading effect from water condensates. 

Panel (f) shows the corresponding entrainment rate from the surrounding atmosphere. It is important to note the updraft begins at a pressure of around 5 bars, with a moist static energy of 4175 kJ/kg. Large entrainment leads to the updraft having a lower static energy at the cloud top, as shown by the parcels that reach a pressure of ~4 bars (light green lines in Figure~\ref{fig:1d_updraft_profile}. Smaller entrainment allows the parcel to retain its relatively high moist static energy, and this is the case for those updrafts that reach an altitude about 1 bar (dark blue lines). The small increase in $\lambda$ at about 1300 hPa corresponds to the decrease in static energy (panel b) at the same altitude. Larger entrainment is required to achieve neutral buoyancy at the local minima.

The cloud condensates within the updraft is shown in the panel (d). Due to the cold temperatures, the cloud particles within the updraft are pure ice. The vertical cloud particle profile within the updraft is maintained by two major processes. The first is from vapor saturation within the updraft, (i.e., the $d(\eta q)$ term). The second is from entrainment of moisture (the $\qut d\eta$ term, where $\qut$ is the total mass mixing ratio of all phases for a condensing species). For most of the deep convecting clouds, which reach high altitudes, the first term dominates, since $\lambda$ is small. This results in condensates within the updraft being much denser compared to the stratiform clouds in the environment. 

\begin{figure*}[ht]
	\includegraphics[width=0.8\textwidth]{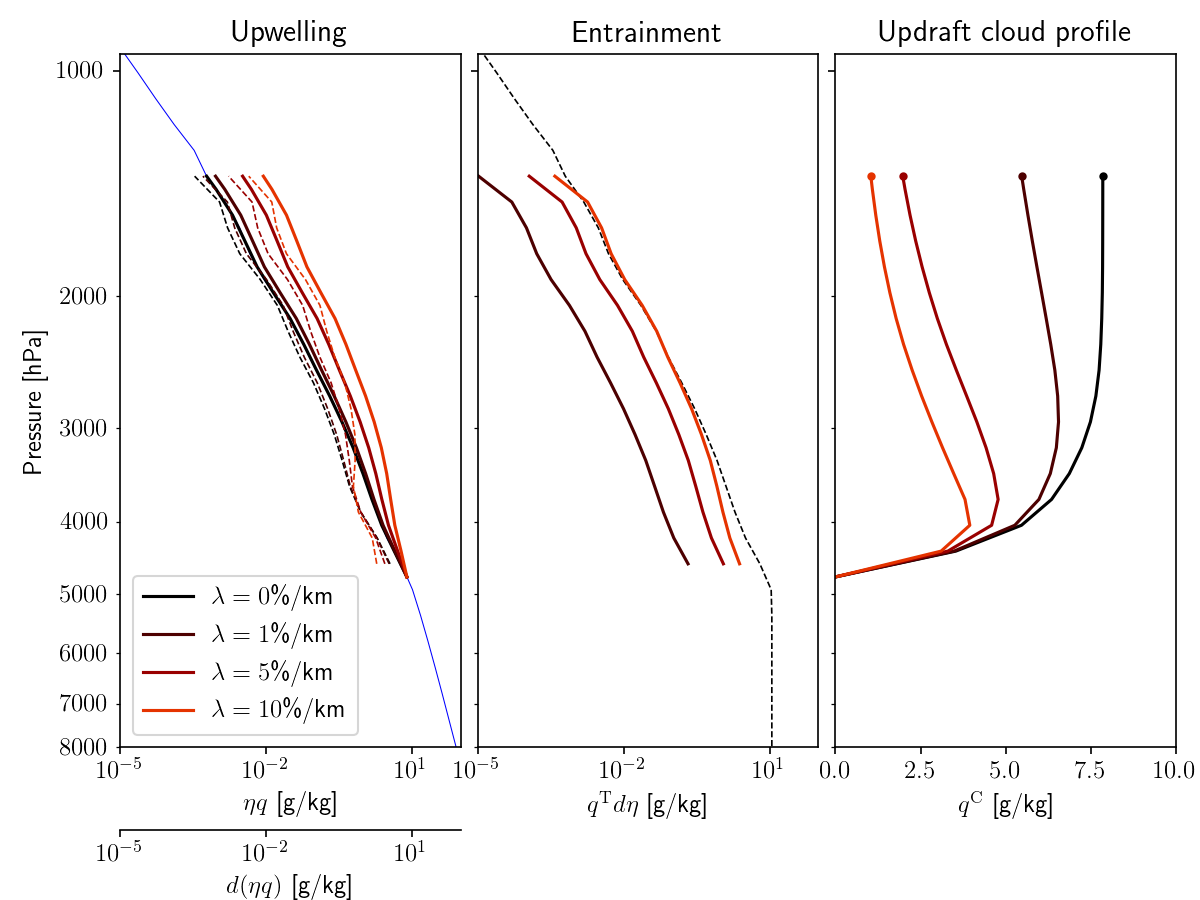}
	\caption{Cloud updraft profiles showing the combined effect of upwelling and entrainment. Panel (a) shows the contribution from vertical advection, with the solid line showing $\eta q$, the dashed lines showing $d(\eta q)$ and the thin blue line showing the saturation specific humidity. Panel (b) shows the contribution from entrainment, with the thin dashed line showing $q^{\rm T}$ and the solid lines showing $q^{\rm T} d\eta$. Panel (c) shows the cloud updraft profile. In all cases, the derivates are taken with respect to the layer number, and are thus dimensionless. All horizontal axes are in [g/kg].
	}
	\label{fig:updraft_cloud_comparison}
\end{figure*}

Figure~\ref{fig:updraft_cloud_comparison} shows the effect of entrainment on these two parameters. Panel (a) shows the  contribution from upwelling and panel (b) shows the contribution from entrainment. For low entrainment values, the vertical mass flux profile is small, and thus the contribution from upwelling is low, and the vertical $q$ profile follows the saturation vapor curve (blue in the panel a). As entrainment increases, so does the vertical mass flux, but also the contribution from entrainment. Therefore, while both processes increase the vertical mass flux of cloud condensate, entrainment generally brings in drier air from the surroundings, while only the upwelling retains moisture rich air. As entrainment increases, the updraft cloud profile decreases (panel c of Figure~\ref{fig:updraft_cloud_comparison}). In the limit of $\lambda=0$, the rate of change of the vertical cloud mass flux profile is equal to rate of change of the vertical vapor flux profile. Consequently, as the vapor is advected upwards, it is limited by the saturation vapor specific humidity at that pressure, and excess vapor within the updraft (i.e., over the saturation limit) condenses to form cloud particles. 
Therefore, the vertical cloud profile is effectively the difference between the updraft vapor profile and the environmental
saturation vapor profile.

\subsection{2D test cases: sensitivity to water abundance}
\label{sec:2dsims}
In the RAS scheme, convection is triggered by an increase in CAPE due to large scale effects (i.e. the grid scale dynamics). Therefore, testing the convective trigger in our model is difficult in the 1D scenario, since there is very little grid scale motion. Therefore, we run a suite of 2D test cases, with the goal of validating the trigger for convection in the RAS scheme, and to test the hyperparameters that control the intensity of convection
, such as the amount of water in the atmosphere. As stated in Section~\ref{sec:intro}, the `mass-loading' effect of water due to its higher molar mass in the H/He atmosphere leads to a negative feedback in the convective activity. Theoretically, with a global abundance of greater than $5\times$ solar [O/H], the convective activity due to water should vanish \citep{Guillot1995,Li2015,Leconte2017}. We test this limit in our simulations, by varying our water abundance between 0.5 to 10$\times$ the solar [O/H] value. Since ammonia generally does not play a role in convective upwelling, we maintain
the value of ammonia at $2\times$ solar, consistent with recent analysis by \citet{Li2017}. 

Preliminary tests showed that the numerical stability of the atmosphere is strongly tied to 
the initial relative humidity of the atmosphere. For higher relative humidity, cloud formation is really fast during the first few hours which destabilizes the model. Above 70\% initial relative humidity, 
the higher abundances are generally unstable. Below this initial relative humidity, 
there is no noticeable difference in the model. Therefore, we start our simulations with 70\% 
initial relative humidity for both water and ammonia.

Moist convection in our model is a response to large scale forcing (the convective mass flux is directly given by the changes in the large scale CAPE). Therefore, we follow \citet{Sugiyama2014}, and use their method for forcing moist convective upwelling in the jovian atmosphere. We apply a cooling of $0.01$ K/day in the region between 100-2000 hPa, which corresponds to the observed upper level cooling by the Galileo probe \citep{Sromovsky1998}. We also apply a heating of 0.003 K/day, between 10000-20000 hPa, which results in a net energy difference of $\sim7.5$ W/m$^2$, which is equivalent to the internal heat flux released from Jupiter \citep{Li2018}. We run our simulations for 200 days to observe the effect of the decrease in upper level stability of the atmosphere and the effect of moist convection in restoring balance. We limit it to this time so that we observe only the effects of moist convection. Over time, the cooling results in an adjustment from large scale dynamics, which makes it difficult to distinguish between adjustment from the large (dry convection) and sub grid scale processes (moist convection). For this reason, we run a simulation without moist convection, and stop our analysis after the moist convective and the `stratiform' cases are similar. Table \ref{tab:2dcases} show the list of simulations. 

\begin{table}[h]
    \centering
    \begin{tabular}{c c}
        \hline
        Case & Water abundance  \\
         & [solar] \\
        \hline
        II.1 & 0.5\\
        II.2 & 1\\
        II.3 & 2\\
        II.4 & 4\\
        II.5 & 5\\
        II.6 & 6\\
        II.7 & 10\\
    \end{tabular}
    \caption{List of cases for the 2D water abundances simulations}
    \label{tab:2dcases}
\end{table}

\begin{figure} 
    \centering
	\includegraphics[width=\columnwidth]{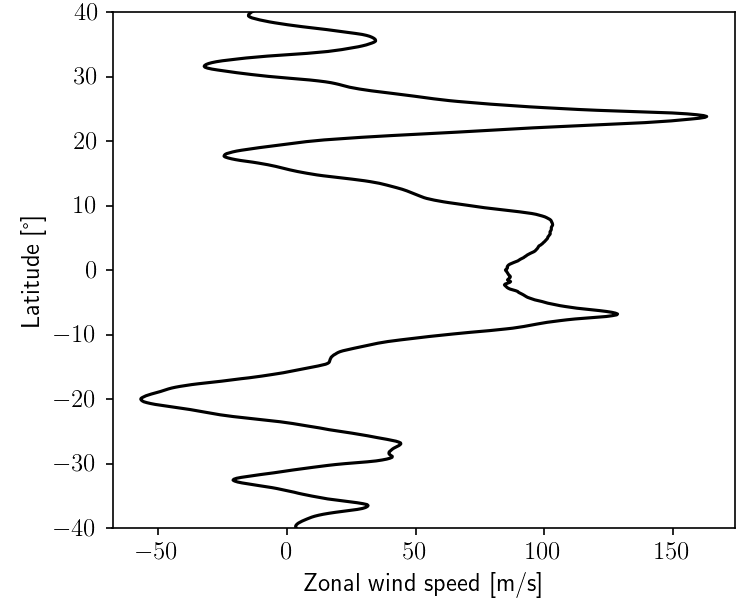}
    \caption{Zonal wind profile used for the 2D test cases. }
    \label{fig:2d_zonalwind}
\end{figure}

\begin{figure} 
    \centering
	\includegraphics[width=\columnwidth]{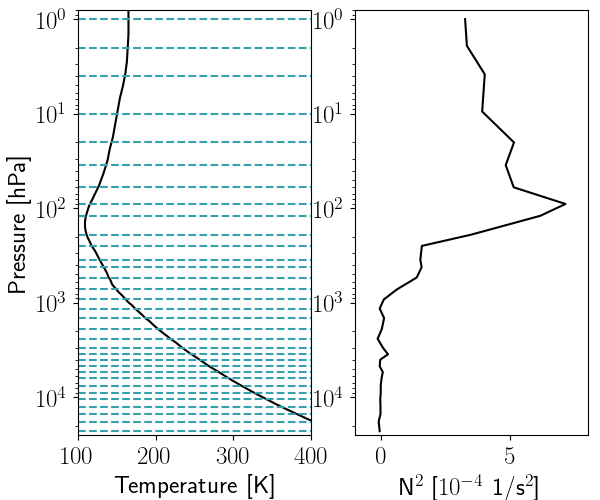}
    \caption{Left: Temperature-pressure of the atmosphere. The dashed blue lines denote model layers. Right: Static stability (given by the square of the Brunt-V{\"a}is{\"a}l{\"a} frequency) of the model atmosphere. Both profiles are for the lower latitude boundary.  }
    \label{fig:tp_bv}
\end{figure}

Our 2D model extends from $40\degree$ S to $40\degree$ N latitude with 200 grid points spaced equally. This gives a resolution of $0.4\degree$ or about 500 km. The initial zonal wind profile for our model is from \citet{Limaye1986} and is shown in Figure~\ref{fig:2d_zonalwind}. We make the assumption of no vertical wind shear with depth below $680$ hPa, similar to the results of \citet{SanchezLavega2008,SanchezLavega2017}. Above this, the wind decays to zero in $2.4$ scale heights as determined by \citet{Gierasch1986}, similar to \citet{Palotai2014}.

Vertically, we cover pressure ranges from $0.1$ hPa down to $22800$ hPa with 32 layers. The layers are placed non-uniformly and are higher resolution near regions of interest (such as the water and ammonia clouds). We add a 4-layer sponge at the top of the model (i.e., $p<10$ hPa) to dissipate vertically propogating waves, and apply Rayleigh drag to the bottom 4 layers ($p>12500$ hPa) to maintain the strength of the deep winds. Neither of these processes are in the region of interest (the weather layer is between 200-6000 hPa, where the clouds form); they are merely to improve model stability at the vertical boundaries. We also apply 8th order hyperviscosity and divergence damping to reduce high frequency modes in the model. The coefficients for these are given by, 
$\nu_8 = (0.5)\left[(1/2400) \Delta y^8/\Delta t\right]$ m$^8$/s, and $\nu_{\rm div} = (0.5)\left[(1/30)\Delta y^2/\Delta t\right]$ m$^2$/s. We place the transition from potential temperature to pressure at 20 hPa, which is above the tropopause, so that this transition in coordinates (i.e., the `seam') doesn't interfere with the tropospheric dynamics. The vertical layer profile, the initial temperature profile from \citet{Moses2005}, and the corresponding static stability is shown in the right panel of Figure~\ref{fig:tp_bv}. 

\subsubsection{Cloud phenomenology}
\begin{figure*}[ht]
	\includegraphics[width=\textwidth]{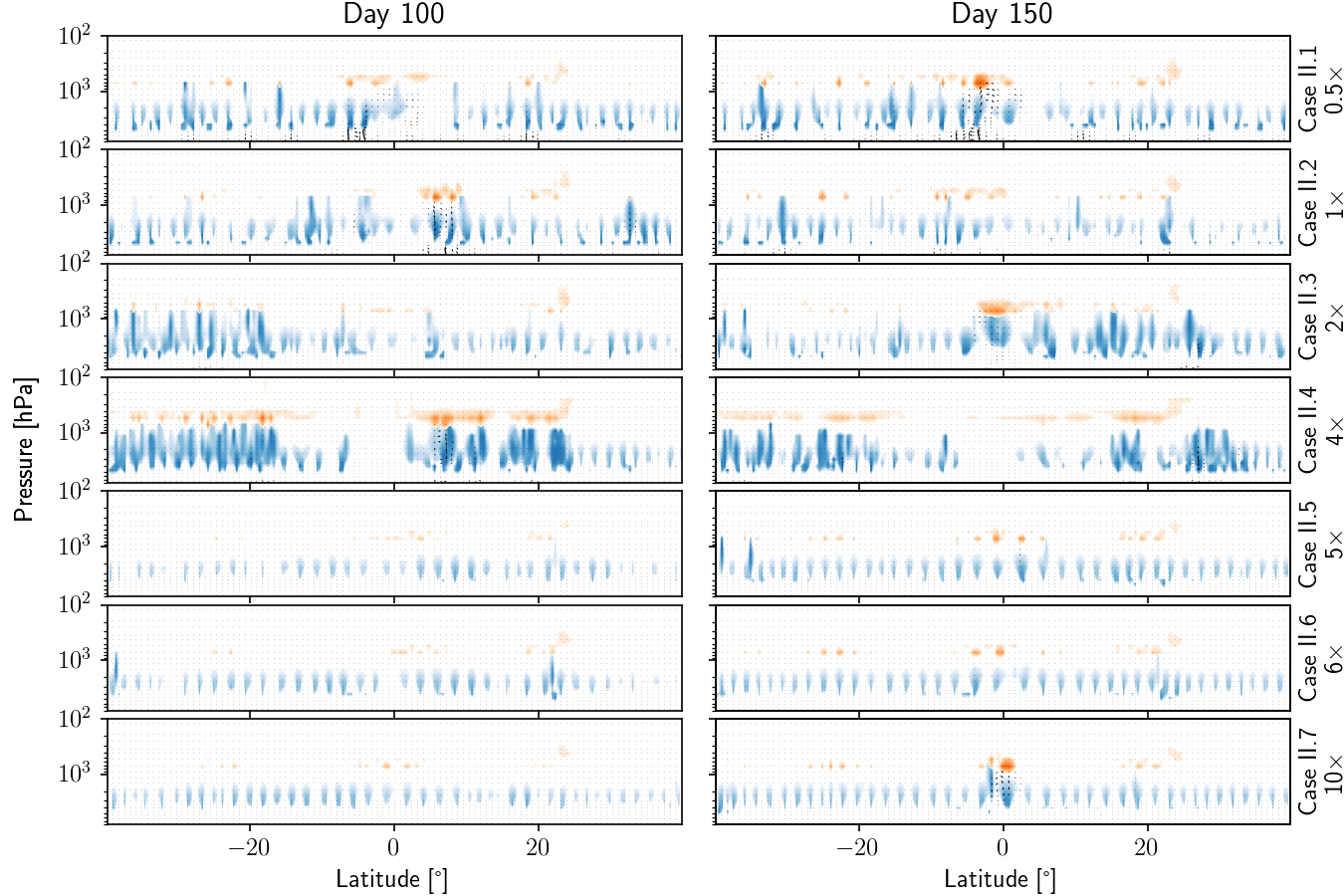}
	\caption{Water (blue) and ammonia (orange) cloud density at day 100 (left) and 150 (right) for different water abundances.}
	\label{fig:cloud_abundance}
\end{figure*}
Fig~\ref{fig:cloud_abundance} shows the cloud phenomonology for the different abundance cases
100 and 150 days into the simulation. The water cloud densities are shown in blue and the ammonia clouds are
shown in orange. 
Cases II.1-4 show large scale cloud formation. The clouds form predominantly near the equator, south of the 24$\degree$ N jet and near $30\degree$ S, and they evolve between the two times. For all cases,
the cloud formations are similar in terms of size and location. The thickest, and tallest water clouds 
reach the base of the ammonia layer, and are driven by moist convective motion rather than large scale
dynamics. The water storms are also accompanied by upper level ammonia clouds. 
The largest clouds are in the equatorial zone and in the belt just south of the $20\degree$ N latitude. There are some 
storms that form at around $20\degree$ S. 

Cases II.5-7 show little cloud activity throughout the model. Both cases show
short clouds dispersed throughout the region which persist at both times without much difference. 
At day 150, Case II.7 shows a storm forming near the equator. However, this storm is still driven by
large scale motion, as indicated by the wind vectors in its vicinity. Therefore, even here,
this is moreso a dynamical response to the instability in the atmosphere (i.e dry convection), rather than one driven by moist convection (which would not require large scale vertical motion). 

\subsubsection{Convection and storm formation}

\begin{figure}[ht]
	\includegraphics[width=\columnwidth]{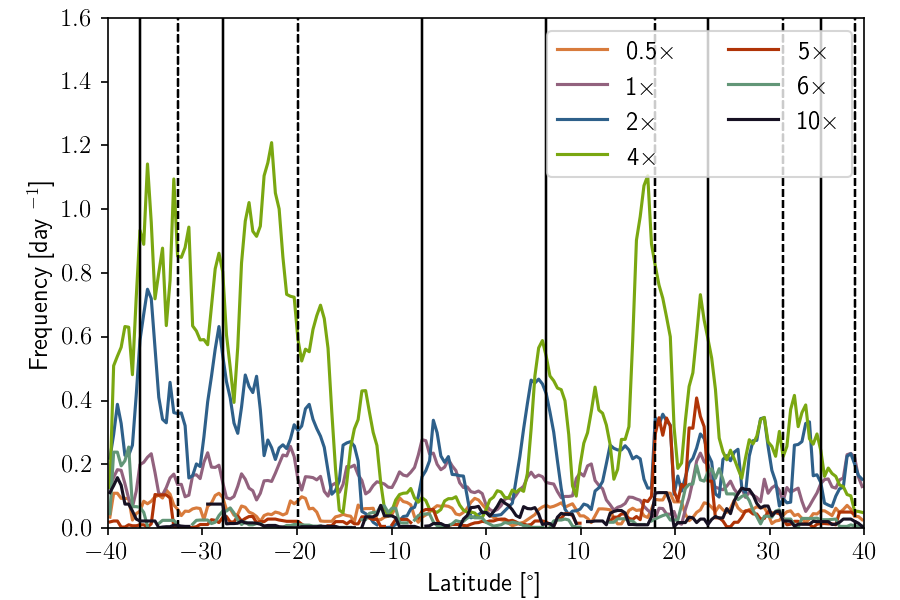}
	\caption{Distribution of storm frequency for different values of water abundance. The solid black lines are locations of prograde jet peaks, while the dashed lines are retrograde jet peaks. }
	\label{fig:storm_lat_abundance}
\end{figure}
\begin{figure}[ht]
	\includegraphics[width=\columnwidth]{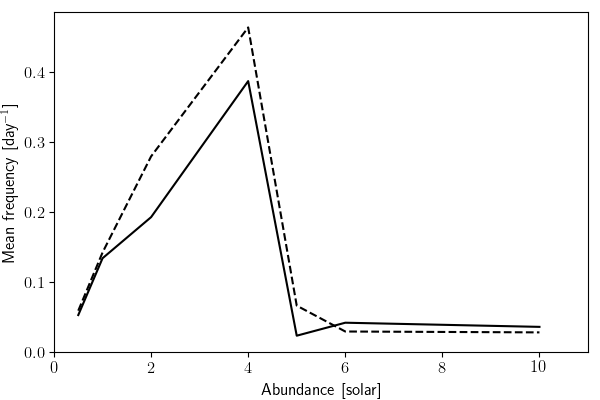}
	\caption{Distribution of storm frequency for different values of water abundance. 
		The values have been averaged for each latitude point over the duration
		of the simulation.}
	\label{fig:storm_abundance}
\end{figure}

Fig~\ref{fig:storm_lat_abundance} shows the distribution of the storm frequency as a function
of latitude for the different abundances. For cases II.2-4, the storm formation is generally distributed throughout the domain. However, there are also specific regions where storm formation is 
noticeably higher. These include the region around $30\degree$ S, between $5\degree-20\degree$ N 
latitude and near the peak of the $24\degree$ N jet. These regions are more prominent in 
Cases II.3 and II.4 compared to Case II.2, with the regions around $30\degree$ S and near $24\degree$ N showing the most convective activity. For cases II.1, 5, 6, 7, there are no noticeable regions that display strong convective potential. As expected, the moist convective events follow regions of dynamical instabilities, as a lot of convective peaks occur near the jet peaks (which are on the boundary between cyclonic and anticyclonic shear, by definition), since sub grid scale moist convection is forced by large scale instabilities. 
Fig~\ref{fig:storm_abundance} shows the mean distribution of storms for different water abundances. This clearly shows the drop-off in convective activity above $5-6\times$ solar water abundance is evident here, as a result of the `mass-loading' effect of the heavy condensibles.

The distribution of moist convective activity for cases II.2-4 compare favorably with 
the detection of lightning using the {\it Juno} Microwave Radiometer (MWR) data \citep{Brown2018}, 
which shows peaks in lightning detection around $20\degree$ S, and between $10\degree-25\degree$ N.
Our results, particularly match the equatorial peak at ${\sim} 8\degree$ N latitude. We also observe less convective activity near the equatorial region for all cases. However, there are differences between the observed convective profile our model results, which we attribute to the following:

\subsubsection{Volatile distribution}
\citet{Brown2018} suggest that the discrepancy in belt/zone moist convection might be due
to variation in the water abundance in those regions. Indeed, \citet{Li2017} shows that this is true for the distribution of ammonia, which is highly variable, even below $1$ bar, at different latitudes. In our model, we assume that the deep water abundance is constant at different latitudes, which would explain the nearly constant meridional distribution.

Furthermore, the equatorial region contains a significantly higher concentration of ammonia vapor 
\citep{Li2017}, compared to the rest of the planet. In this region, the increased ammonia concentration
would lead to higher atmospheric molar mass, compared to other regions. This would consequently lead to 
a lower value for CAPE, since the parcel density would be closer to the atmospheric density here, compared
to one that is drier. 

\subsubsection{Forcing profile}
We assume a constant forcing as a function of latitude, which is likely true, if we 
account only for radiation. However, moist convective activity is also driven 
meridional circulation cells, vorticity, etc., which cannot be simulated effectively in this
2-dimensional setting. \citet{Brown2018} also suggest the increase in moist convective activity 
above $50\degree$ latitude is suggestive of preferentially poleward distribution of the internal heat flux. 
	
\subsubsection{Response of the dynamics}
\label{sec:dynamics_response}
\begin{figure}[ht]
	\includegraphics[width=\columnwidth]{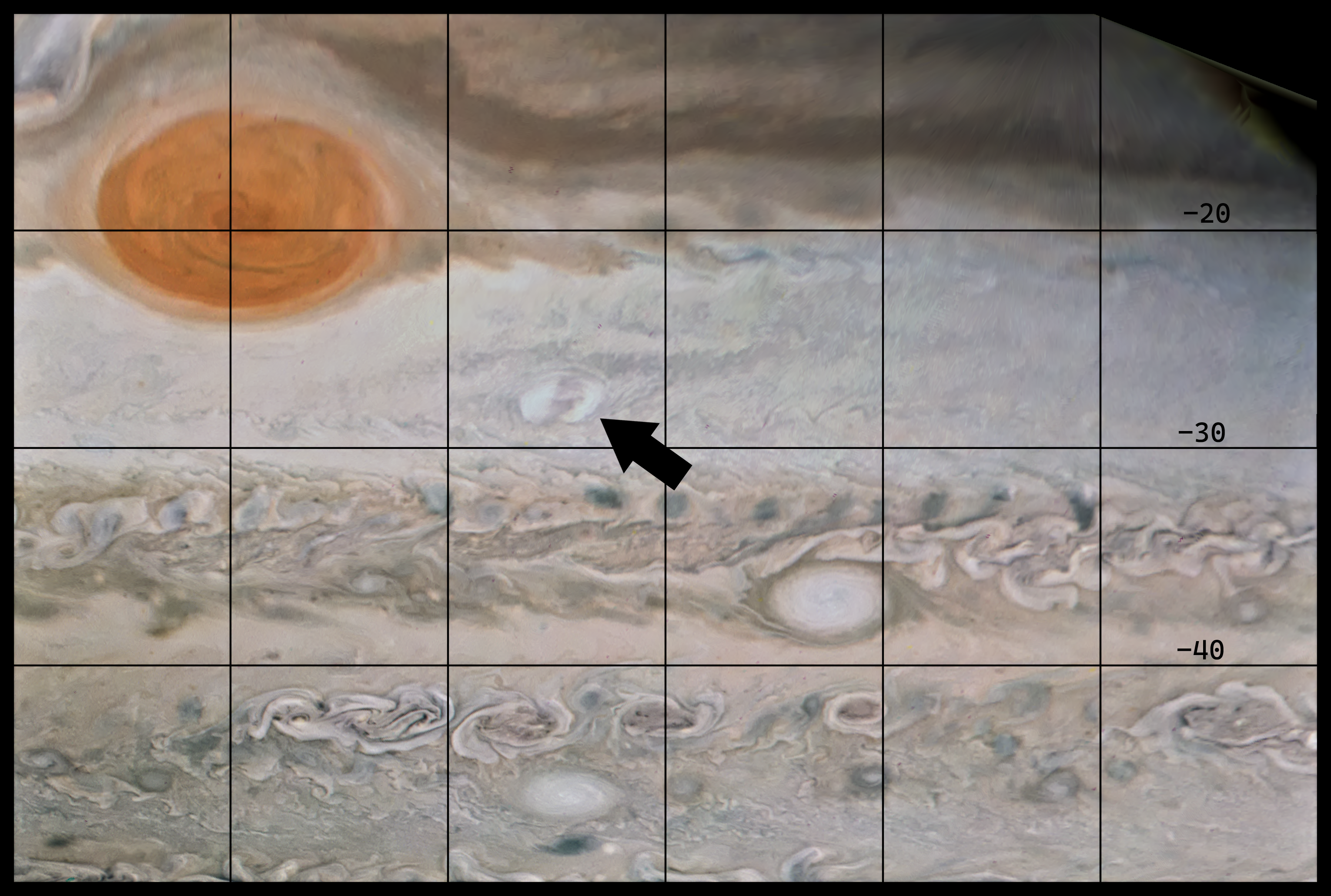}
	\caption{{\it JunoCam} image of the region between $10\degree-50\degree$ S. The evolution of 
	Clyde's spot, a moist convective upheaval is visible to the south-east of the Great Red Spot (black arrow). }
	\label{fig:GRS_junocam}
\end{figure}
The large scale dynamics also act as a response to the increasing upper level instability. Therefore, 
the response to the forcing is split partially into the response from water convection and, partly from
the changes in the atmospheric structure from large scale dynamics. 
The increased convective activity at $30\degree$ S in our model is particularly striking,
compared to what was determined by the MWR data. The region around $30\degree$ S shows 
moist convective activity across all the cases, implying that this is indeed a feature of the model
that seemingly does not agree with the natural conditions on Jupiter. Fig~\ref{fig:GRS_junocam}
shows the region from {\it Juno's} JunoCam, with the latitude values annotated on the right side. The 
region between $30\degree$ and $40\degree$ S are filled with turbulent cyclonic features, suggesting
that dynamical instability are a common feature in this latitudinal band. However, the lack of lightning
detections suggests that moist convective triggers are minimal. 
Therefore, it is likely that the instabilities manifest quickly as cyclonic vortices, thereby reducing,
or completely negating the need for a moist convective outburst. This is indeed similar to the findings 
of \citet{Showman2007}, who found that adding energy to the atmosphere at higher latitudes resulted 
in the emergence of vortices, rather than jets. He found that the critical latitude for the transition between
jets and vortices was around $45\degree$, which is comparable to the latitude of this region. 

Alternatively, storm formation in the region near the $30\degree$ S latitude might not be strong enough to generate lightning, or not have 
the proper atmospheric structure to generate detectible lightning storms. This region might promote
shallow convection, due to the response from the dynamics reducing the instability in the upper layer. 
It should be noted, however, that there are likely detections of lightning below a pressure level of 
5 bars \citep{Vasavada2005}, where the temperatures are too warm for ice to form. The liquid clouds are
fairly inefficient in generating sufficient charge separation for lightning to form \citep{Gibbard1995}.
Therefore, these observations are currently unexplained.  

\subsection{Summary of test cases}
In summary, we have tested our moist convective scheme in both the 1D and the 2D setting. In the 1D test cases, we validated the vertical profiles of the sub grid-scale convection, and ensured that the buoyancy and energy constraints were being met. 

In the 2D tests, we forced the atmosphere to become unstable, thereby triggering moist convection. We tested our model with different water abundances to verify the sensitivity of the convective intensity to the amount of water in the atmosphere. We found that our moist convective scheme validates the mass loading effect of water, whereby above a water mixing ratio greater than $5\times$ the solar value, there is a stark decrease in the amount of convection, since the water in the atmosphere creates a stabilizing layer against convection. We also found that moist convection in our model responds well to dynamical forcing (as evidenced by the increase in moist convection near regions of dynamical instability, such as the jet peaks). 

As described in Section~\ref{sec:mcscheme}, there are a two major unknowns that the new scheme is sensitive to: the relaxation timescale ($\tau$), and the trigger mechanism (currently determined through finite difference, as the rate of change of atmospheric CAPE). The relaxation timescale used in our model ($\tau\sim 1-2$ hours) is comparable to modeling studies by \citet{Hueso2001}, who find that that individual storms develop over the timescale of a few hours. The trigger mechanism is generally much more difficult to determine. In Earth models, convection is triggered by comparing the model value of CAPE to a critical value determined from atmospheric soundings \citep{ArakawaSchubert1974}. This is not possible on Jupiter, and this mechanism will need to be investigated further in future works, perhaps through models that directly resolve convection \citep{Hueso2001}, or treat the atmosphere as a fully compressible fluid \citep{Ge2020}. With the current implementation, we find from our 2D test cases that the model generally responds as expected to atmospheric forcing, but perhaps a much more robust implementation is required for matching all the observations, such as the lack of convection near $30\degree$ S. Therefore, with the current implementation, we treat these as numerical parameters that need to be tested for sensitivity. 

Furthermore, since these tests were run in the 2D framework, we find that it is difficult to generalize the convective nature of the jovian atmosphere from this analysis, since there are a lot of convective events that have a predominantly 3D structure, e.g., STB Ghost \citep{Inurrigarro2020}, or Clyde's Spot, as shown in Figure~\ref{fig:GRS_junocam}. These require further detailed study with a fully 3D model, that capture their respective physics accurately, as stated in Section~\ref{sec:dynamics_response}, in the same vein as Section~\ref{sec:3dsim} below. 

\section{The 24$\degree$ N jet }
\label{sec:3dsim}
With the new moist convective scheme validated, we move to three dimensional test cases. The goal here is to study the coupling between the large scale dynamics and the moist convective storms. Furthermore, we focus on the region surrounding the $24\degree$ N jet to study the origins of the observed moist convective plumes, and their effect on, and response to, the atmospheric flow. In the 3D setting, we can fully resolve the growth and propagation of sub grid scale convective storms. Our goal in this study is to investigate the aftermath of the initial plume outbreak, similarly to \citetalias{Sankar2021}. The periodic nature of the initial plume hints at complex origin \citep{Fletcher2017b}, which we do not model here as that represents a unique state of the atmosphere. Instead, we are interested in understanding the general convective nature of this region, which is well depicted in the aftermath of the outbreak.  We further study the role of water in fueling these storms, and its effect on the atmospheric flow, and structure of the jet. \citetalias{Sankar2021} found that the cloud formation and convection in this region was driven by the formation of an upper level baroclinic potential vorticity (PV) wave. The cyclonic wave packet demonstrated increased convective potential, while the anti-cyclonic packet diminished convection. Our goal here is to extend that study with simulations containing a full parameterization of convection, in order to resolve the convective intensity directly. 

\subsection{Model setup}

\begin{figure}
	\centering
	\includegraphics[width=\columnwidth]{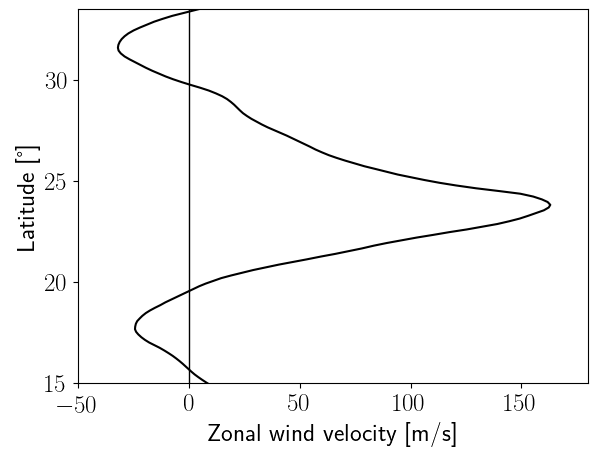}
    \caption{Zonal wind profile used to initialize the model. This wind profile  applied at a pressure of $680$ hPa in our model assuming no wind shear below and wind decay over 2.4 scale-height above.}
    \label{fig:zonal_wind}
\end{figure}

\subsubsection{Grid parameters}
Our model grid for the 3D cases is the same as the one in \citetalias{Sankar2021} and covers the region from 15$\degree$-$33.5\degree$ N  with 80 grid points in the meridional direction. Zonally, the model covers $0\degree$-$120\degree$ longitude, with 256 grid points. This produces a resolution of $0.47\degree$ per grid point zonally and $0.23\degree$ per grid point meridionally. The angular resolution is higher in the meridional direction to compensate for the geometric effect at the mid-latitudes, ensuring that the physical extents of the grid points are roughly equal (about $200$ km per side, in this case). 
The zonal wind profile used is from {\it Voyager} data from \citet{Limaye1986} and is shown in Figure~\ref{fig:zonal_wind} for the model region. We use the same vertical wind shear assumption as in our 2D test cases, and our vertical temperature structure is the same as the 2D cases described above, and is shown in Figure~\ref{fig:tp_bv}. 

\subsubsection{Model initialization}
One of the main goals of this study is to interpret the effects of different water abundances on the resulting atmospheric and cloud structure, and on the convective capability of the atmosphere. To compare the new convective parameterization with previous EPIC simulations of this region without the RAS scheme, we use the same initialization as in \citetalias{Sankar2021}, and test $0.5\times$, $1\times$ and $2\times$ solar for water. Given the lack of difference between the ammonia abundance on both the convective potential and the dynamics of the wave, we maintain the deep ammonia abundance at $2\times$ the solar [N/H] ratio, which is consistent with \citet{Li2017}.

Similarly to \citetalias{Sankar2021}, we also allow the model to evolve for 200 days without any cloud processes enabled, in order to let the model adjust to the initial parameters. This is to ensure that there are  no unphysical cloud formation in the first few timsteps, which affect model results. We also use the same perturbation profile to remove the zonal symmetry set up in our model as a result of this `spin up' phase, whereby we induce 100 small vortices (up to 10 m/s in speed) randomly throughout the model between a pressure of 500-5000 hPa. We use three different set of these random vortices to ensure that our model results is not dependent on the profile of the perturbation. We test three different profiles for each deep abundance value of water. The list of test cases is shown in Table~\ref{tab:cases}.

\begin{table}
	\caption{List of cases for the 3D tests. The perturbation column corresponds to the different random vortex profiles used to remove zonal symmetry from the model (as detailed in the text)}
	\label{tab:cases}
	\begin{center}
		\begin{tabular}{ccc}
			\hline
			Case & H$_2$O   & Perturbation \\
				 &  [solar] & \\
			\hline
			III.1  & 0.5 & 1 \\
			III.2  & 1   & 1 \\
			III.3  & 2   & 1 \\
			III.4  & 0.5 & 2 \\
			III.5  & 1   & 2 \\
			III.6  & 2   & 2 \\
			III.7  & 0.5 & 3 \\
			III.8  & 1   & 3 \\
			III.9  & 2   & 3 \\
		\end{tabular}
	\end{center}
\end{table}

\subsubsection{RAS parameters}
For the RAS scheme, we tested simulations with $\tau=30$ min, 
$\tau=1$ hour and $\tau=2$ hours for all cases. We noticed that the simulations with $\tau = 30$  mins produced excessive horizontal and vertical cloud coverage that was inconsistent with observations. We did not observe this effect in the simulations with $\tau=1$ and $\tau=2$ hours. 
Therefore, for the following analysis, we will only use $\tau=1$ hour for the $0.5\times$ and $1\times$ cases and $\tau=2$ hours for the $2\times$ cases. The different relaxation times lead to the same result towards the end of the simulation, but the higher relaxation times for the $2\times$ case allows for better stability in the beginning. 

With the moist convective scheme active, the atmosphere quickly evolves into a similar wave-driven structure seen in the stratiform case \citepalias{Sankar2021}. The dynamics of the motion of the wave results in up- and downwelling that leads to distinct cloud features. There are arc shaped ammonia clouds and deep water clouds that are extremely localized due to the same circulation described in \citetalias{Sankar2021}. In the sections below, we look at the differences between the stratiform cases and the addition of the new convective module, to investigate the effect of moist convection in this region. We also draw new conclusions on the convective potential of this region, and implications for the water abundance. 

\subsection{Results}
Similarly to \citetalias{Sankar2021}, we found that cloud formation and convection is driven by the formation of an upper level potential vorticity wave in our model. The cyclonic wave packet produces thicker clouds, while the anticyclonic packet produces cloud clearing.

\subsubsection{Dynamics of the upper level wave}
\begin{figure}
    \centering
    \includegraphics[width=\columnwidth]{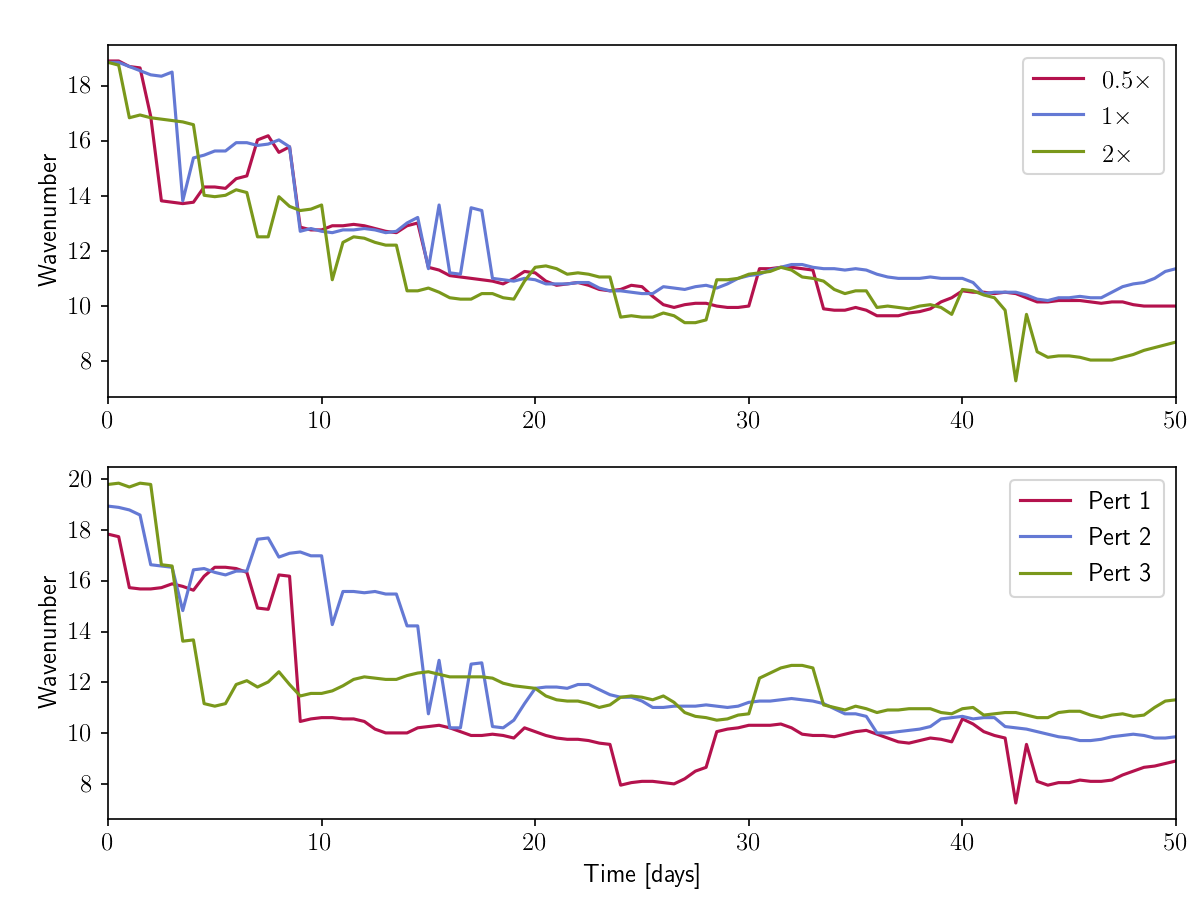}
	\caption{Evolution of wavenumber as a function of time, for different water abundances (top) and perturbation profiles (bottom).}
	\label{fig:moist_f_vs_t}
\end{figure}
Figure~\ref{fig:moist_f_vs_t} shows the change in the zonal wavenumber of the wave with time. The trend for the different cases is downward with time (i.e. the wave stretches zonally with time), similar to the stratiform case detailed in \citetalias{Sankar2021}. The $0.5\times$ and $1\times$ stabilize at a wavenumber of $\sim 10$, while the $2\times$ case drops to a wavenumber of $\sim 8$ after about 40 days. The stratiform case reaches this state a few days later.

Overall, the moist convective upwellings do not affect the wave, and the dissipation (i.e., lengthening of the wave over time) happens at roughly the same pace as before. Understandably, this is likely due to the moist convective plumes reaching a peak well below the location of the wave, and thus is a negligible change to the thermodynamic structure of the upper atmosphere. 

\subsubsection{Ammonia cloud morphology}
\begin{figure*}
    \centering
    \includegraphics[width=\textwidth]{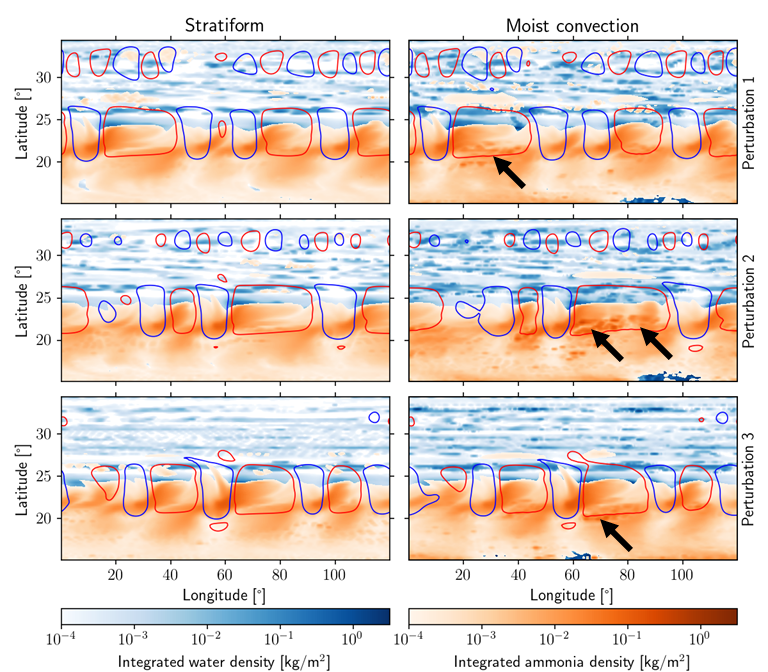}
	\caption{Difference between stratiform (left) and convective (right) cases for $2\times$ solar for different perturbation profiles. Integrated cloud densities for ammonia clouds are shown in orange and water in blue. The upper level wave is shown with the red/blue contour, with red corresponding to cyclonic eddy vorticity and blue corresponding to anticyclonic eddy vorticity, both showing the $\pm 0.4$ PVU or $\pm 4\times10^{-7}$ K m$^2$/kg s isosurface. The black arrows point to locations of moist convective storms, as a consequence of turning on the RAS scheme.}
	\label{fig:cloud_diff_h2}
\end{figure*}
Figure~\ref{fig:cloud_diff_h2} shows the difference between the stratiform and convective cloud morphology for the $2\times$ solar cases (Cases III.3, 6 and 9) at day 45. The blue and red contour show the location of the wave as determined by calculating the eddy potential vorticity (i.e., the departure of the Ertel potential vorticity from the zonal mean value). The red contour corresponds to regions that have a cyclonic (positive) PV anomaly and the blue corresponds to anticyclonic (negative) PV anomaly. There is very little difference in the large scale structure of the ammonia clouds within the jet. The arc shape is still present, and the dynamics of the ammonia cloud match that of the stratiform cases. Understandably, this is due to the fact that ammonia does not form convective clouds in the model, and the effect of moist convection on ammonia would be through the upwelling of vapor from the depth, with the water-induced storms. 

\begin{figure*}
    \centering
    \includegraphics[width=\textwidth]{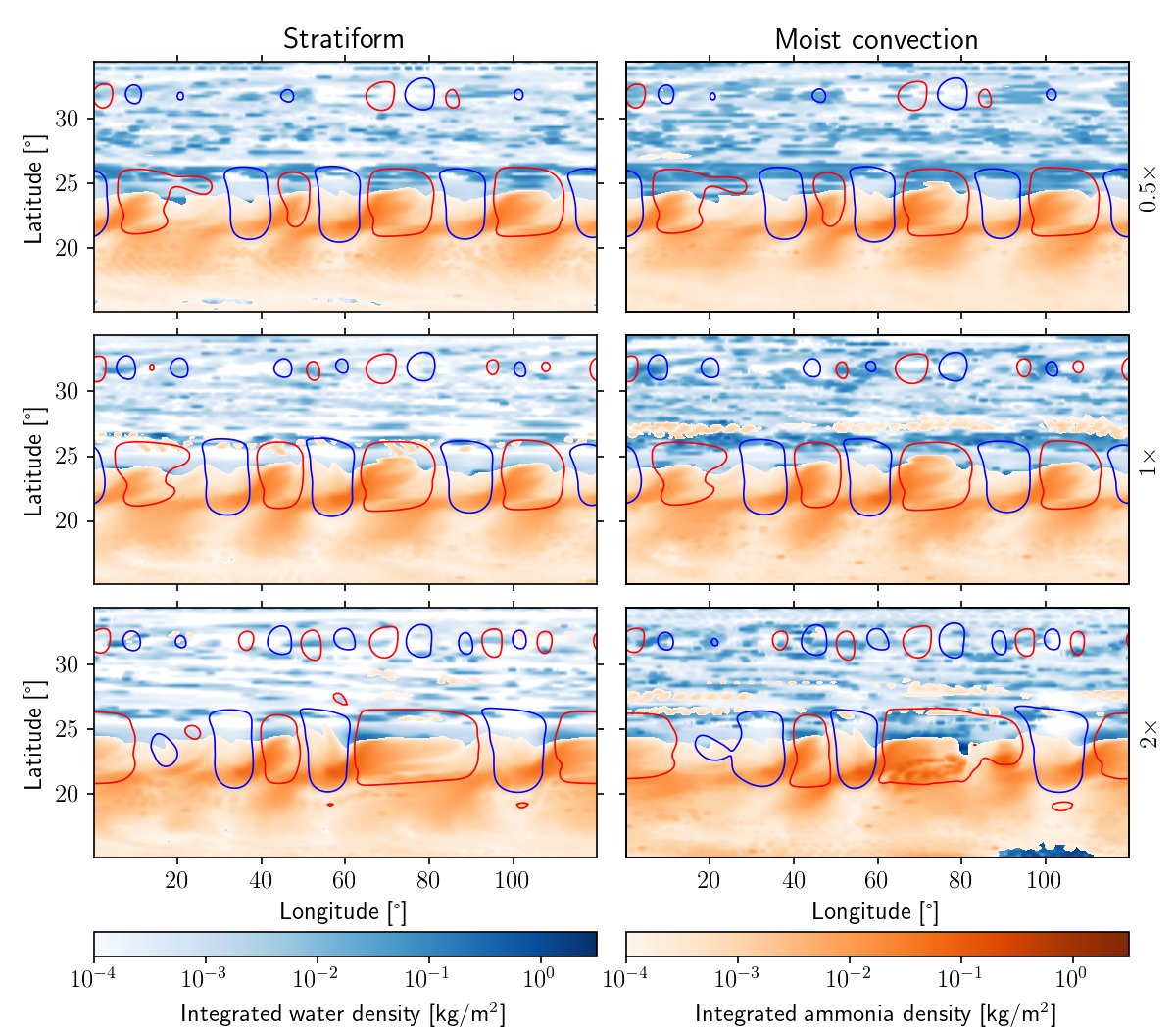}
	\caption{Differences between stratiform (left) and convective (right) cases for Perturbation 2, as a function of water abundance. The PV contours in red/blue show the upper level wave, similarly to Figure~\ref{fig:cloud_diff_h2}.}
	\label{fig:cloud_diff_abundance}
\end{figure*}
Indeed, the moist convective cases show an abundance of small scale outbursts, highlighted by the black arrows in Figure~\ref{fig:cloud_diff_h2}. These outbursts result in thick ammonia clouds as the moist convection brings ammonia rich vapor to the upper atmosphere, and are primarily concentrated to the south of the jet. They do disrupt the structure of the arc shapes within the wave, but are very short-lived, lasting on the order of less than a day. 
The number of such small scale moist convective updrafts increases with water abundances, as shown in Figure~\ref{fig:cloud_diff_abundance}. This is expected, as the driver for convection is the water.  

\subsubsection{Water cloud morphology}

\begin{figure}
    \centering
    \includegraphics[width=\columnwidth]{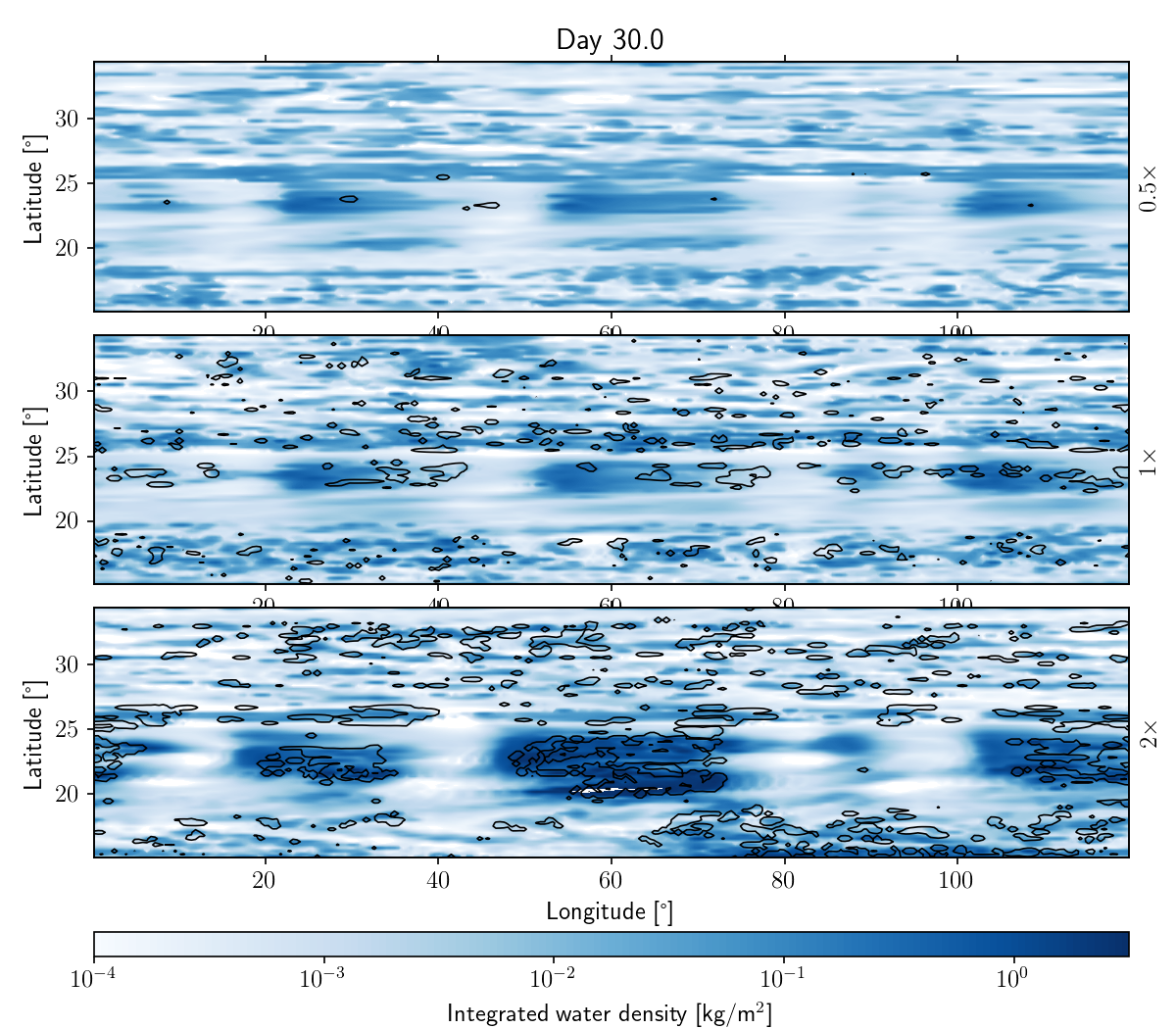}
	\caption{Integrated water cloud density in blue and water mass flux in black contours for different water abundances, for perturbation 1, 30 days into the convective simulation. The black contours denote regions of strong water convection. }
	\label{fig:water_mass_flux}
\end{figure}
Figure~\ref{fig:water_mass_flux} shows the water cloud density for perturbation 1 at day 30 for different water abundances, along with the convective mass flux in black. The mass flux, and the cloud density increase with the abundance of water, as expected. Furthermore, the storm formation is generally constrained to specific zonal bands, and is the highest near the jet peak. The storms within the jet lead the cyclonic packet of the wave (i.e. are on the eastern side of the cyclonic packets). The location of the wave, and the effect on the cloud density is consistent between the stratiform and convective cases, and also between the different water abundances. 

\begin{figure}
    \centering
    \includegraphics[width=\columnwidth]{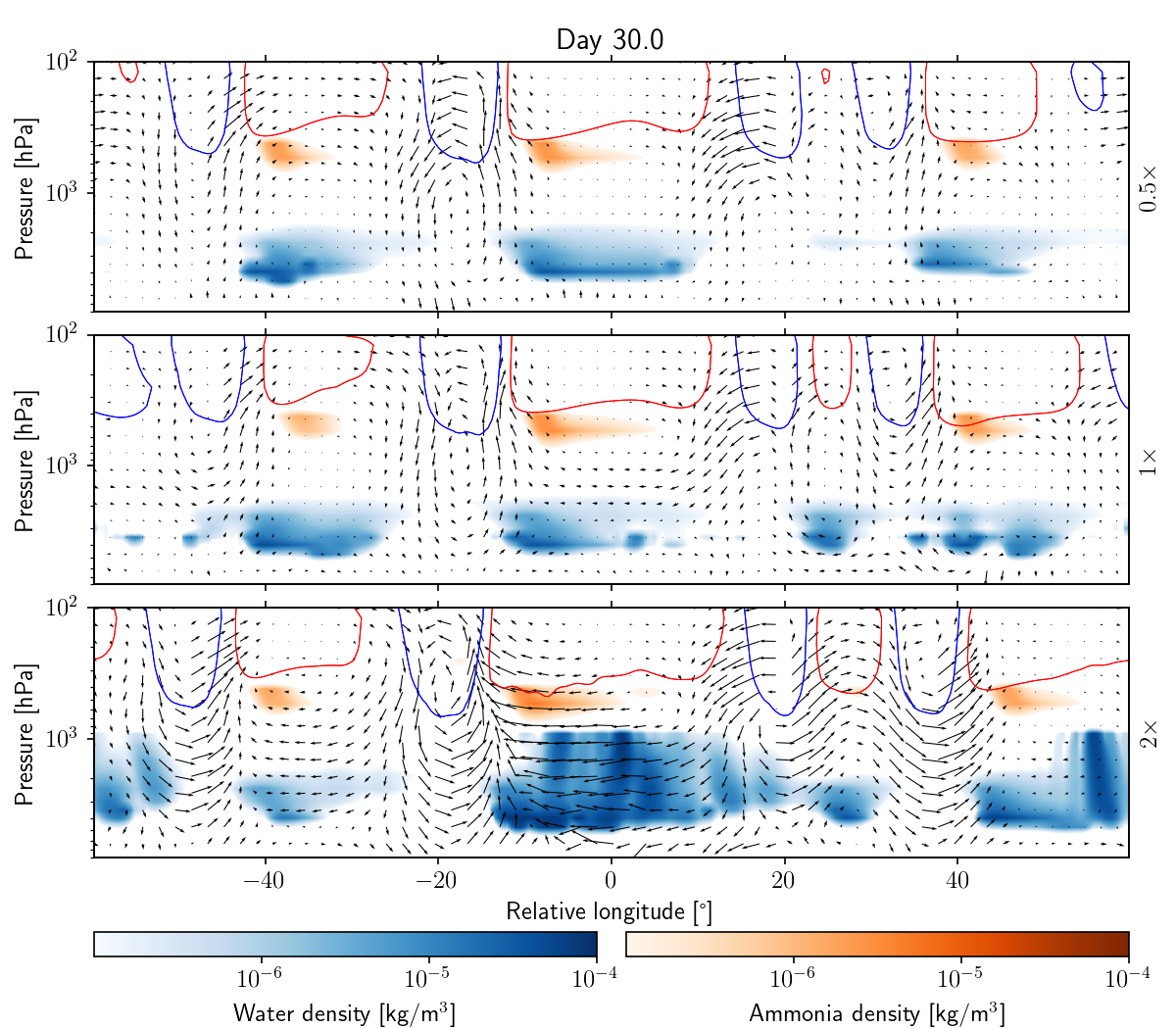}
	\caption{Zonal slice across the jet peak centered on the cyclonic part of the wave, showing the ammonia density in orange and the water density in blue, for different water abundances at day 30. The cyclonic packet is shown with the red line, and the anticyclonic packet is shown in blue, both corresponding to the 0.4 PVU isosurface as in Figure~\ref{fig:cloud_diff_h2}. The vertical velocities have been scaled up by a factor of 20 to enhance visiblity. In this frame, the maximum upward velocities measure about 30 cm/s}
	\label{fig:moist_slice_frame_day30}
\end{figure}

Figure~\ref{fig:moist_slice_frame_day30} shows the zonal slice across the jet peak for perturbation 1, 
for different water abundances, 30 days into the convective simulations. We see the same cloud morphology as the stratiform cases, with the thicker clouds existing below the cyclonic part of the wave, beginning to the east of the upwelling, while the anticyclonic part of the wave is still generally clear, and has downwelling. The cloud density corresponds directly to the amount of water, with the $2\times$ case
having the thickest clouds. 

The upwelling drives water clouds up to the 1-bar level, in the $2\times$ cases, which changes the habit for water ice, both in terms of the air density and ambient temperature, compared to the water cloud base. This changes the microphysical profile of the water ice particles within these updrafted water clouds, which is the key difference between simulations with, and without, the RAS scheme. 

\begin{figure}
    \centering
    \includegraphics[width=\columnwidth]{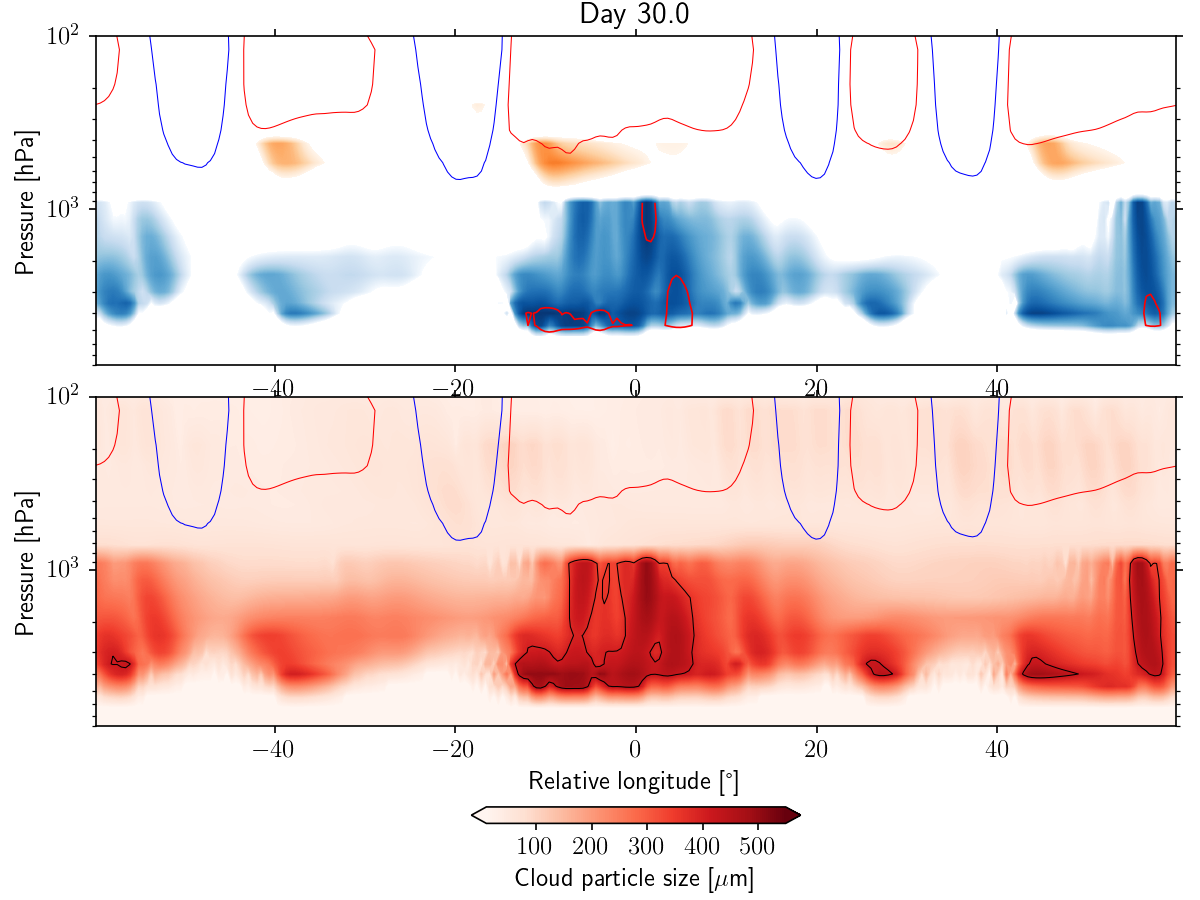}
	\caption{Microphysical parameters for the water clouds for Case III.3, 20 days into the simulation. Top: Ammonia cloud density in orange and water cloud density in blue (using the same scale as Figure~\ref{fig:moist_slice_frame_day30}), with the cyclonic and anticylonic wavepackets plotted in thin red and blue contours for the 0.4 PVU isosurface. The thick red line shows the locations where water snow exceeds a density of 10 g/m$^3$. Bottom: Ice cloud particle size (in $\mu$m) plotted with the filled contours. The black line denotes the isosurface of ice cloud particles falling with terminal velocities greater than 50 cm/s.}
	\label{fig:micro_pert1_day_20}
\end{figure}

Figure~\ref{fig:micro_pert1_day_20} shows the cloud density for water and ammonia clouds in the top panel and cloud particle size in the bottom panel 30 days into the simulation with the first perturbation profile. Particles falling with terminal velocity greater than 50 cm/s are shown in black. As expected, particle sizes are largest within the upwelling, and thus have the largest terminal velocities. The red contours show the location of snow particles, which occurs where particle sizes exceed $500~\mu$m. Snow forms at the top of the upwelling and fall with the downdraft. Therefore, both the precipitation and the downwelling deplete the updrafted cloud particles on the eastern edge of the wave, which returns the water mass back to the deeper levels. 

Within the updraft, the particle sizes are generally between 300-500 $\mu$m for the ice clouds. The fastest falling particles correspond to those just at the critical diameter of $500~\mu$m and fall with a terminal velocity of 70 cm/s, at a pressure level of 800-900 hPa. As these particles fall, their terminal velocity drops to about 58  cm/s near the cloud base at 4 bars. At an average velocity of 65 cm/s, this journey takes about 16 hours from the top of the storm (900 hPa) to the cloud base -- a total height of about 37 km. 

While snow particles are larger, their larger surface area decreases the terminal velocity compared to the bullet-shaped cloud ice particles. Therefore, while snow particles reach diameters up to 1 mm, near the cloud base, the terminal velocity of snow within a grid cell reaches at most 85-90 cm/s, which is not significantly higher than that of cloud ice. 

Horizontally, the distance from the eastern edge of the cyclonic packet to the western edge of the next one is about $10\degree$ or about 11000 km at this latitude. At the 1-4 bar level, the horizontal wind speed is about 160 m/s, and thus advecting the water cloud particles from a region of downwelling to the next upwelling takes about 19 hours, which is a little longer than the vertical fall speed of the particles. Therefore, even with the moist convection lofting particles to the upper atmosphere, precipitation still is a viable method of removing volatiles from the convective plume, resulting in a rarefied atmosphere below the anticylonic packet. 

\begin{figure*}
    \centering
    \includegraphics[width=\textwidth]{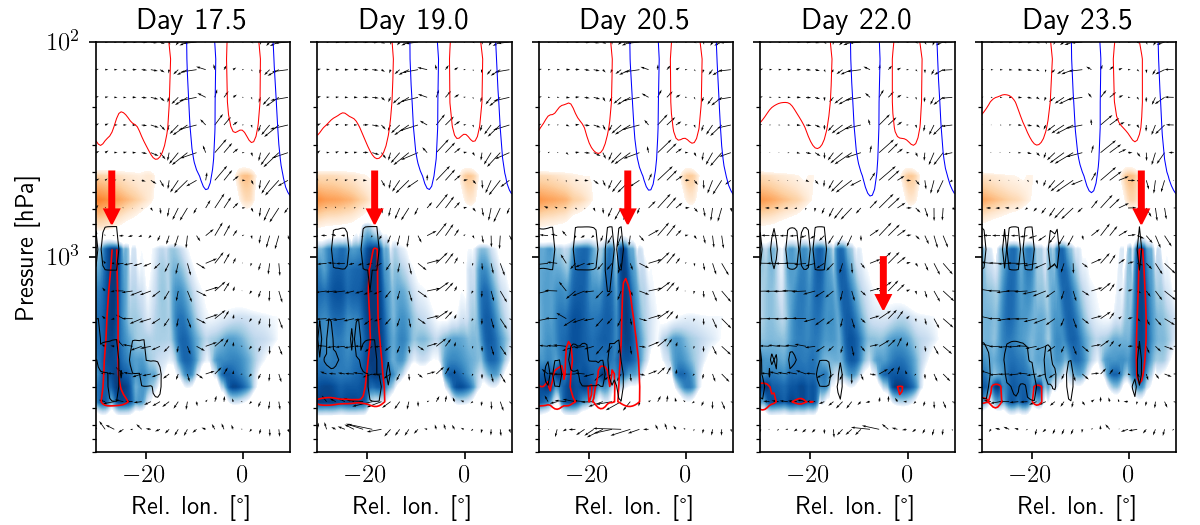}
	\caption{Close-up of the region between two successive cyclonic wave packets, showing the progression of the water cycle. Water cloud density is shown in blue and ammonia in orange. The red contours show the region of snowfall, and the black contours show the tops of moist convective storms. The red arrow follows a water cloud that forms on the left wave packet, disappears and reappears in the next wave, along with the convective mass flux and snowfall. }
	\label{fig:cycle_convection}
\end{figure*}
Therefore, as water clouds travel below the wave, they are lofted up to the east of the upwelling below the cyclonic packet, until they reach the eastern edge, where convection stops, and they sediment to the lower levels. Sedimentation occurs over the anticylonic packet, until they reach the next cyclonic packet, where they undergo moist convective upwelling again, and the process repeats. This cycle is shown in Figure~\ref{fig:cycle_convection}, with the red arrow denoting the packet that demonstrates this effect. As this parcel reaches the eastern edge, the moist convective intensity (shown with the black contours) and the amount of snowfall both decrease, and is restored as it crosses the western edge of the next cyclonic packet. Note how the water cloud dissipates near day 22 as it is pushed downwards, and thus decreases the relative humidity, and consequently, the cloud density. 

\begin{figure}
    \centering
    \includegraphics[width=\columnwidth]{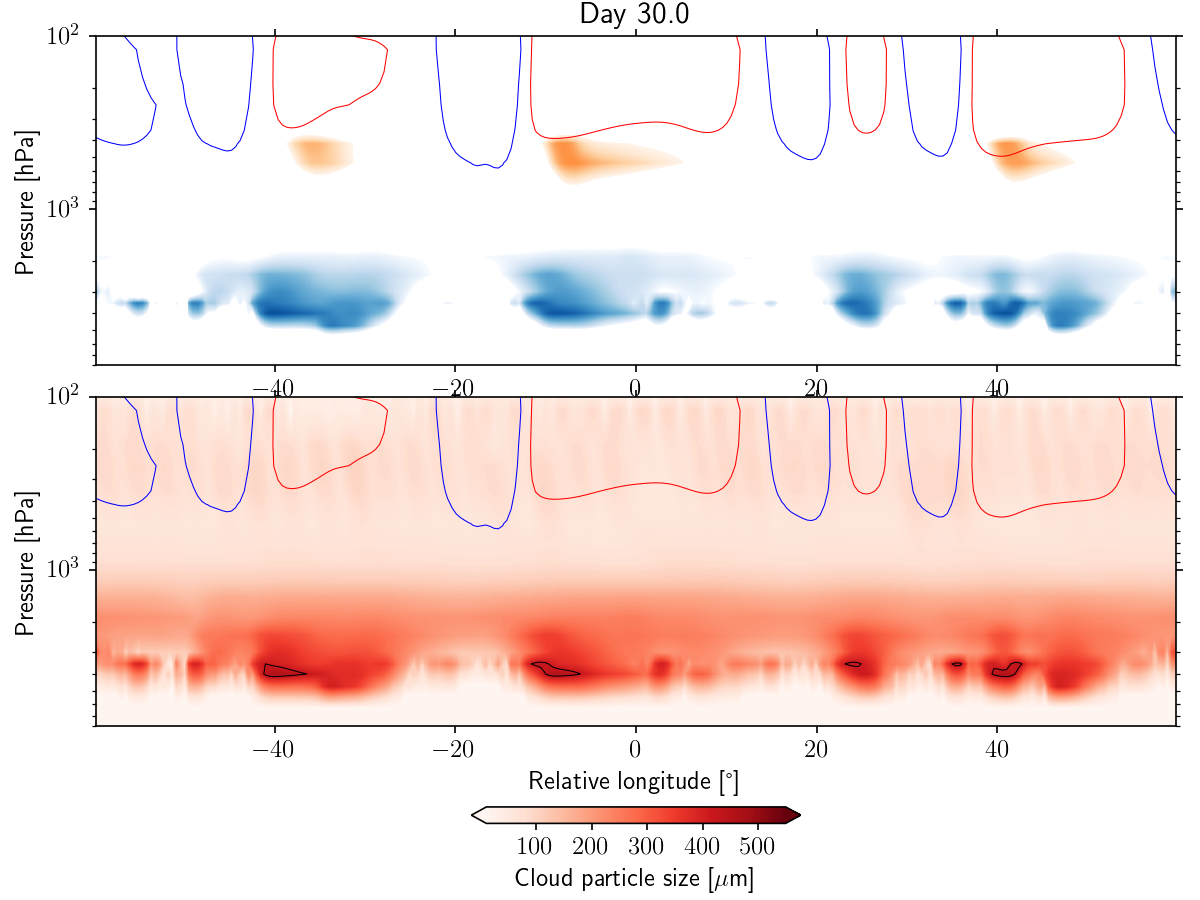}
	\caption{Same as Figure~\ref{fig:micro_pert1_day_20}, but for Case III.2}
	\label{fig:micro_pert1_day_20_h1}
\end{figure}

For the $1\times$ solar case, the microphysical structure of the atmosphere is much more simple (Figure~\ref{fig:micro_pert1_day_20_h1}). The water clouds are shorter, reaching below the 1-2 bar level, and do not form the persistent convective towers in the eastern edge of the cyclonic wave packet. Indeed, both the snow formation and precipitation occur near the cloud base, rather than at 1 bar. Therefore, the water clouds in the $1\times$ cases are much more transient, compared to the $2\times$ case, even with the convective module. 
This is due to the relative lack of convection in the $1\times$ cases, compared to the $2\times$ cases.

\subsubsection{Convection and convective activity}
\begin{figure*}
    \centering
    \includegraphics[width=0.8\textwidth]{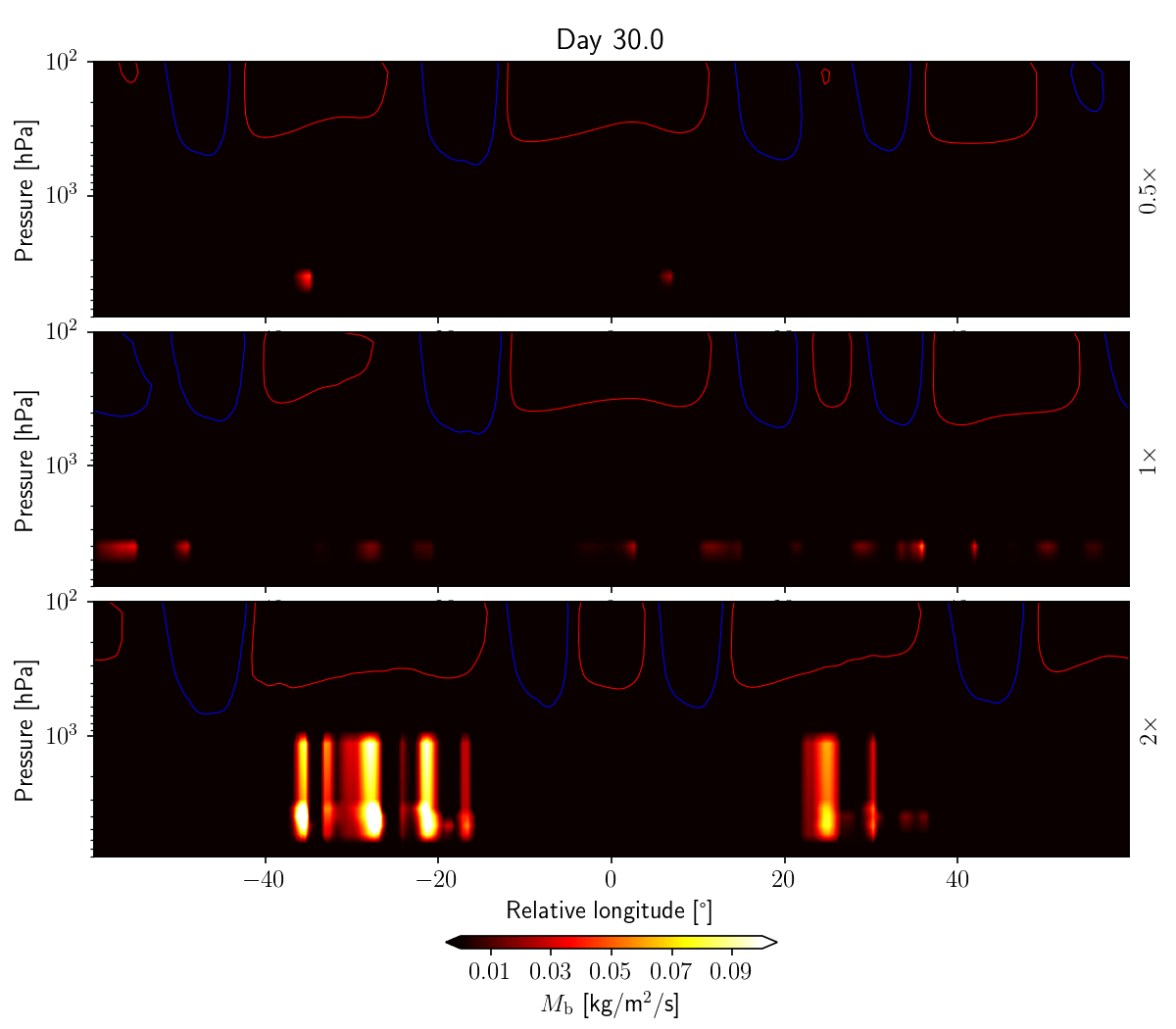}
	\caption{Zonal slice at the jet peak at a latitude of $23.7\degree$ N showing the convective mass flux at day 30 (in kg/m$^2$/s) into the simulation, for different water abundances. The red and blue contours show the location of the cyclonic and anticyclonic wavepackets, respectively, signifying the $\pm 0.4$ PVU isosurface. }
	\label{fig:moist_slice_flux_frame_day30}
\end{figure*}

Figure~\ref{fig:moist_slice_flux_frame_day30} shows the corresponding mass flux profile for the simulations in Figure~\ref{fig:moist_slice_frame_day30}. Here, we can see the increase in convective mass flux directly result in increased cloud density. Only the $2\times$ cases show deep convective storms, while the $1\times$ shows shallow convection. The $0.5\times$ show very little convection. 

For the $2\times$ case, the storms are tied to the location of the convective packet, and more specifically, to the east of the updraft. The convective fluxes stop just to the west of the downwelling. Interestingly, the upwelling itself does not directly result in convection. Indeed, the lack of convection at the western edge of the cyclonic packet actually shows that the upwelling inhibits sub-grid scale convection. Instead, the updrafts adds moisture to the east of its location, which increases the CAPE below the bulk of the cyclonic wave packet, which in turn triggers convection. 

\begin{figure*}
    \centering
    \includegraphics[width=\textwidth]{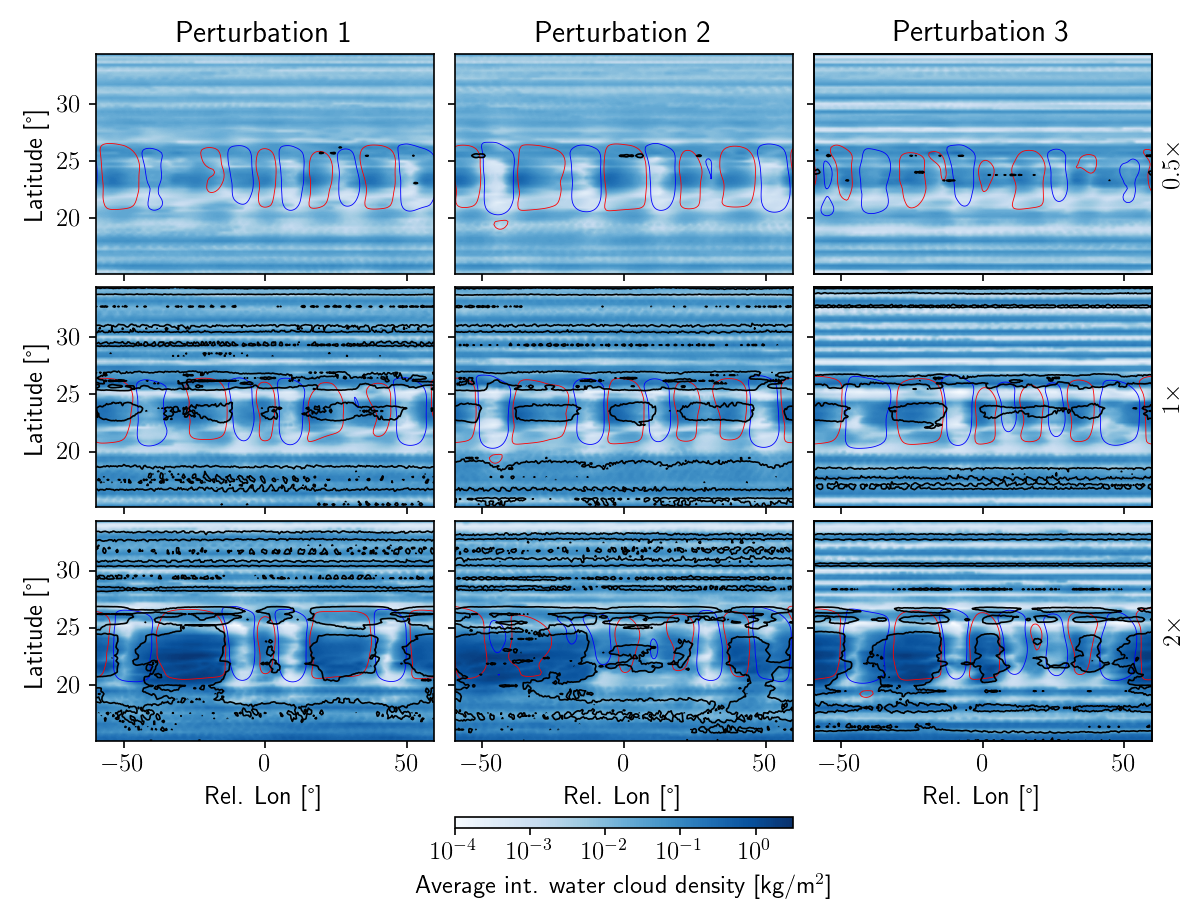}
	\caption{Time averaged water cloud density in blue, in a frame that travels with the wave. The cyclonic and anticyclonic packets are shown in the thin red and blue contours, denoting the positive and negative $0.4$ PVU isosurface, respectively. The black contours show the locations where storm frequencies exceed 0.5 per day. }
	\label{fig:average_nstorm_water}
\end{figure*}

Figure~\ref{fig:average_nstorm_water} shows the average water cloud densities for all nine simulations, and the corresponding location of the wavepackets. The black contours show the locations where storm frequencies exceed 1 every two days. As expected, the $0.5\times$ cases show very little storm frequency and convective mass flux, but the $1\times$ and $2\times$ cases show very strongly localized regions of convection. For the $1\times$ cases, the structure of the wave correlates well with the location of convective events, and the wave itself is maintained well. In the $2\times$ cases, the storms are more disorganized, and bleed into the belt to the south. The wave itself loses its structure from the upwelling (especially in the case of perturbation 2).

\begin{figure*}
    \centering
    \includegraphics[width=\textwidth]{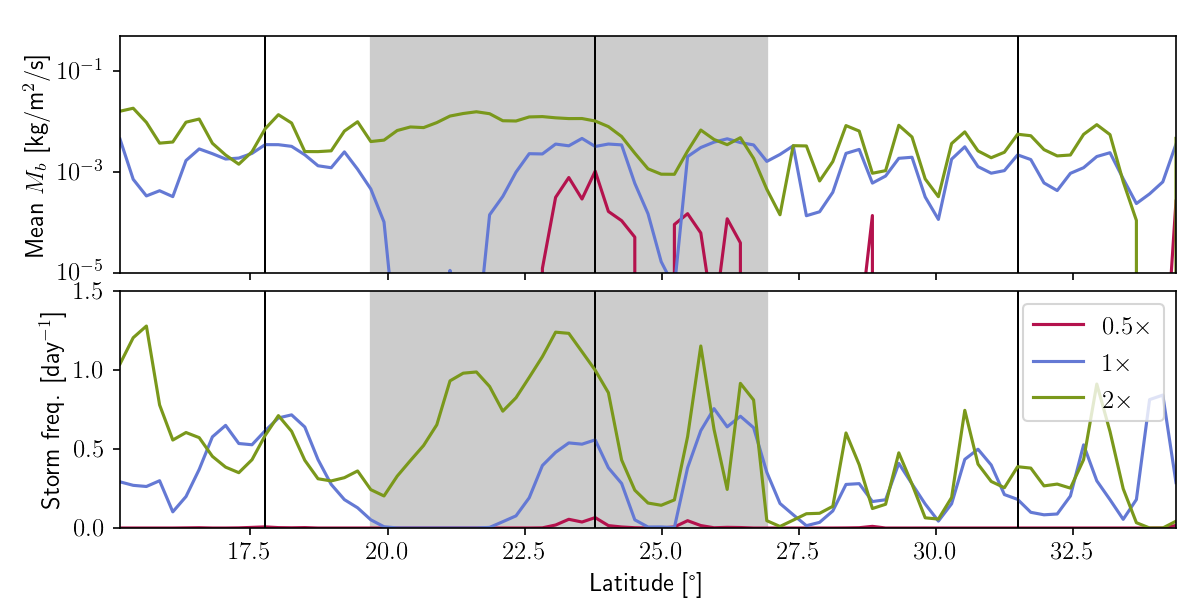}
	\caption{Temporally and zonally averaged mass flux (top) and storm frequency (bottom) for different water abundances. The shaded region corresponds to the $24\degree$ N jet. Vertical black lines denote peaks of the jets in the modeled region. }
	\label{fig:flux_abundance_lat}
\end{figure*}

Figure~\ref{fig:flux_abundance_lat} shows the zonal and temporal mean of the storm frequency and intensity as a function of latitude and abundance. The shaded region denotes the jet, and the vertical black line corresponds to the jet peak. As stated before, the jet itself is the region of highest convective activity across all the cases, and interestingly, does not show uniform distribution of convective storms. Most of the storm formation is to the south, with a decrease in storm formation just to the north of the peak, ending with more storms on the northern edge. Elsewhere, the belt on the southern boundary shows some convective activity, as does the belt to the north. Convection seems to be primarily concentrated near the peaks of the jets at 17.5$\degree$ N and $23.7\degree$ N\@. To the north of $25\degree$ latitude, the convection is distributed throughout, without any distinguishable structure. 

Interestingly, the $2\times$ case show a pretty uniform distribution of storm intensity across the region, even though there is a distribution in storm frequency. Therefore, for the $2\times$ case, the moist convective flux is nearly uniformly everywhere, with differences in the atmospheric structure between the regions resulting in either recurring, low intensity storms, or infrequent, high intensity storms. Near regions of dynamical instability (e.g., at jet peaks), the former is true, while elsewhere, it is the latter. 
The uniformity is likely a function of the initial atmospheric structure of this region, in terms of both the distribution of water and the temperature at depth. 

\begin{figure*}
    \centering
    \includegraphics[width=\textwidth]{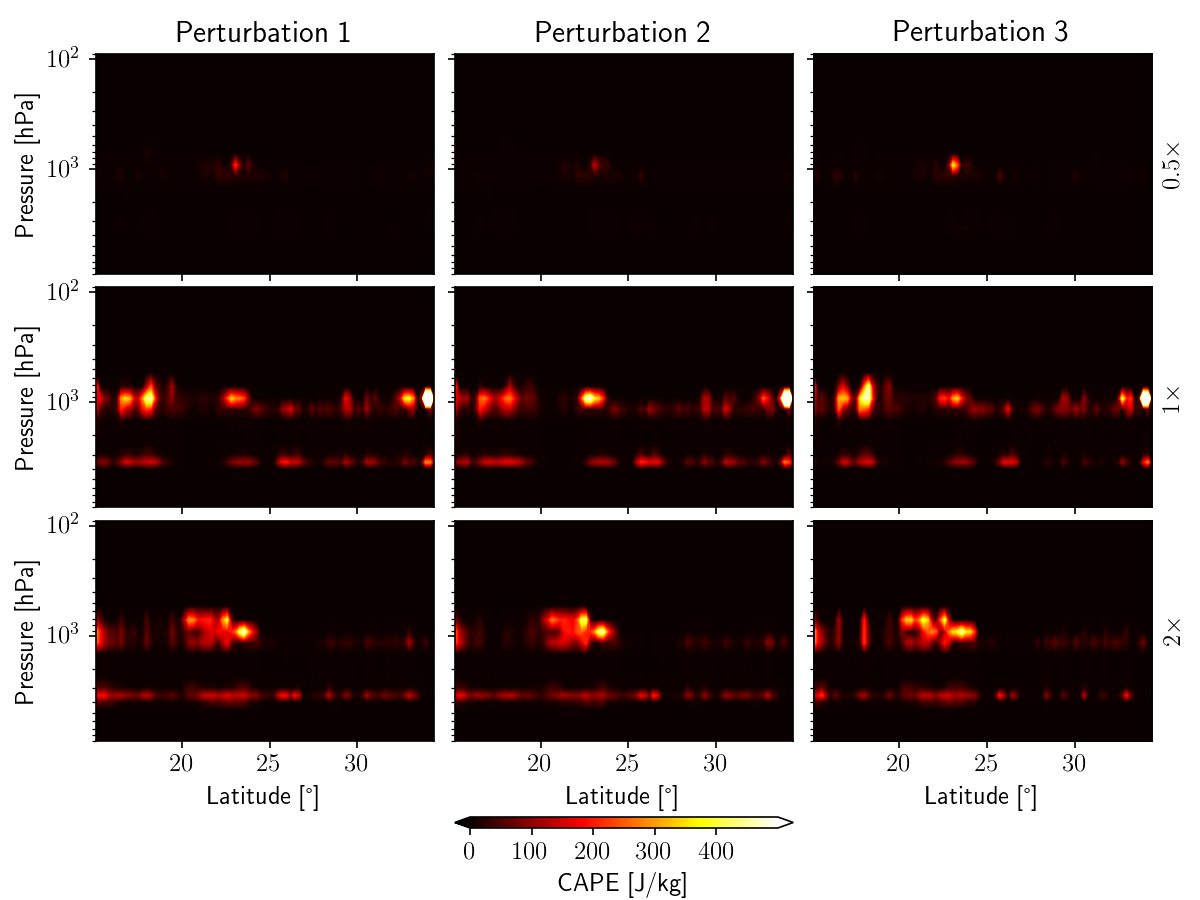}
	\caption{Temporally and zonally averaged value of CAPE for each simulation.  }
	\label{fig:CAPE_lat}
\end{figure*}

The distribution of CAPE, as shown in Figure~\ref{fig:CAPE_lat} reveals a similar profile. Here we show the CAPE determined from the cloud base, following the convective updraft profile up to the top of that cloud type, using Equation~\ref{eq:CAPE}. The $0.5\times$ solar cases show very little CAPE, except at the jet peak. In the $1\times$ and $2\times$ solar cases, the CAPE is concentrated near the jet peak and to the south, which corresponds closely with the distribution of storm formation in Figure~\ref{fig:flux_abundance_lat}. 

The value of CAPE in our model average between 500-1000 J/kg for parcels reaching the 1 bar level, with the higher values corresponding to the larger water abundances. For shallow convection, CAPE was only about 30 J/kg for the $1\times$ cases and about $100$ J/kg for the $2\times$ cases. While this is inconsistent with our stratiform cases, note that CAPE defined here follows the parcels ascent profile such that the convecting cloud is at most neutrally buoyant at the cloud top. Therefore, CAPE defined here also accounts for the decrease in buoyancy from the entrainment of dry material, making it much more consistent with an actual updrafting parcel, and also much smaller than the CAPE determined for the stratiform cases. 

Since CAPE corresponds to the total potential energy gain from convection, we can equate this to the kinetic energy of the parcel at the top of the updraft, and determine an approximate mean vertical velocities within these convective plumes, as,
\begin{equation}
	w_{\rm mean} \sim \dfrac{1}{2}\sqrt{2\times {\rm CAPE}},
\end{equation}

which means that for shallow convection, the mean updraft vertical velocity is under 10 m/s, travelling only about 5-10km. These updrafts thus take less than half an hour to reach the cloud top. With a mean mass flux of about $10^{-3}$ kg/m$^2$/s (Figure~\ref{fig:flux_abundance_lat}), which is approximately equal to the base mass flux, since these clouds only travel 1-2 layers vertically, this means that these shallow updrafts only move a few kilograms of air for every square meter that they cover. 

For the deeper convective cells, reaching the 1 bar level, and a CAPE of about 1000~J/kg, we get a peak updraft vertical velocity of about 45 m/s, and mean velocities of around 20 m/s, which is consistent with moist convective modeling jovian storms \citep{Hueso2001}. Given a vertical height of about 30 km from the cloud base to the top, the updrafts should take under 20 mins, making these storms very efficient at quickly carrying energy upwards. While the cloud base mass flux for these systems are comparable to those for shallow convection, the increase in mass flux, as the cloud entrains, results in a net flux that is an order of magnitude, or more, larger. Note that there are multiple storm cells that reach the layers between 800-2000 hPa, each with a mean cloud base flux of $\sim10^{-3}$ kg/m$^2$/s. Therefore, these deep convective cells transport a correspondingly larger amount of mass and energy vertically, as expected. 

The net energy flux is more complicated to determine. The convection happens as a response to large scale forcing, and so the cloud base flux is created to reduce the increase in CAPE\@. In our simulations, the mean rate of change of CAPE in the deeper levels is only about $10^{-3}$ W/kg, and an order of magnitude larger for the deep convection for the $2\times$ cases. Curiously, the $1\times$ cases show increased CAPE in the 800-1000 hPa region, but very little convection occurs in this region. Indeed, while the CAPE is large, the rate of change of CAPE is nearly negligible, meaning that there is not sufficient forcing in the atmosphere to initiate convection up to the $1000$ hPa level in the $1\times$ cases. 

Intuitively, we can determine the net energy flux observed as,
\begin{equation}
	F \sim \int \rho P_c dz,
\end{equation}
where $P_c$ is the convective energy flux (i.e., rate of change of CAPE), $\rho$ is the atmospheric density and $F$ is the emitted power. Note that this is a trivial, order-of-magnitude estimation, rather than a rigorous calculation. 

\begin{figure*}
    \centering
    \includegraphics[width=\textwidth]{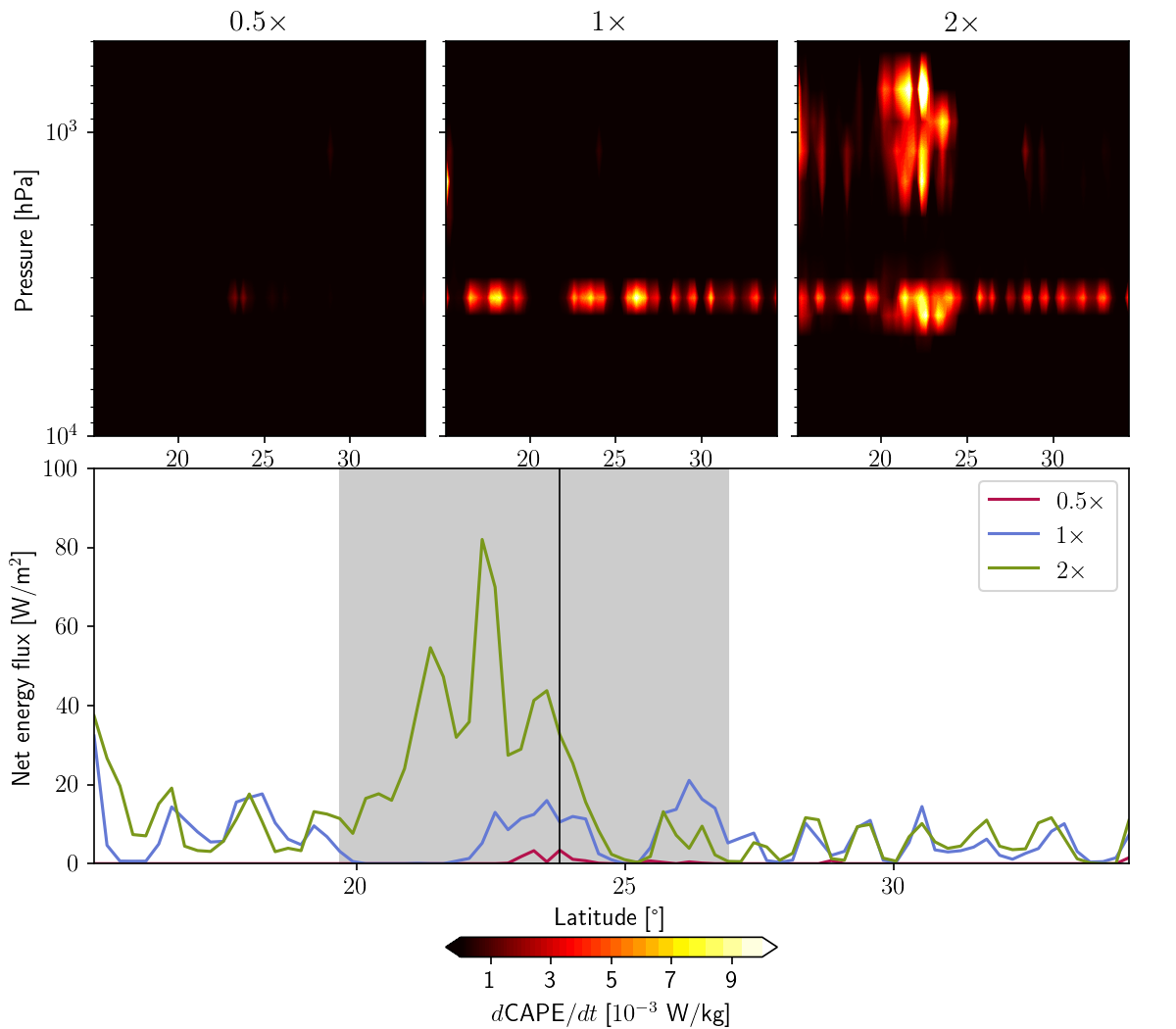}
	\caption{Temporally and zonally averaged value of convective power ($d$CAPE$/dt$, top) for each simulation and the corresponding convective energy flux (bottom). The shaded region denotes the $24\degree$ N jet.}
	\label{fig:flux_lat}
\end{figure*}

Figure~\ref{fig:flux_lat} shows the net energy flux from convective activity as a function of latitude, for different water abundances. Here, we see that the general trends as described above, with the distribution of convection both vertically and meridionally being similar to what was described above. The key difference, however, is in the stark difference in the convective energy flux between the water abundances: for the $0.5\times$ cases, there is very little convective flux, which has a mean value $0.1$ W/m$^2$, the $1\times$ has a mean value of $6.3$ W/m$^2$, while the $2\times$ is around $13$ W/m$^2$. These mean values are represented well in the northern half of the model domain for both the $1\times$ and $2\times$ cases, but near the jet, there is an extremely large increase in the convective energy flux for the $2\times$ case, with a peak of around $80$ W/m$^2$.

\section{Discussion and Conclusions}

In this work, we have detailed the addition of a sub-grid scale moist convective parameterization to the EPIC model and our simulations of convection in the jovian atmosphere, and tested the scheme  to validate our model against the theoretical prediction of moist convection in the jovian atmosphere.
We applied the RAS scheme to study convection and cloud formation near Jupiter's 24$\degree$ N jet. We investigated the effects of varying the deep abundance of water and ammonia and perturbed the atmosphere with random vortices. We found that, similar to our study with a stratiform cloud model \citepalias{Sankar2021}, a wave forms from baroclinic instability and exists above $200$ hPa and the motion of the wave results in up- and downwelling which strongly influences the formation of both water and ammonia clouds. The upwelling occurs on the western edge of the cyclonic anomaly and enriches the upper atmosphere with water vapor, leading to an increase in convective potential and the formation of convective storms. 

For the convective updrafts within the jet, the water clouds formed tall storms that breached the 1-bar level in the $1\times$ cases occasionally, and in the $2\times$ cases consistently. These updrafts were shifted to the east of the large scale updraft within the cyclonic packet, and the downdrafts, and precipitation dissipated the storms below the anticylonic packet. The advection of water clouds below the cyclonic packet clearly showed the formations of storm cells when the advection at the 4 bar level, below the wave, increased the enchriment of water vapor (see Figure~\ref{fig:cycle_convection}). Therefore, these smaller storms would be seen as packets of bright clouds moving against the backdrop of the slower moving wave. 
However, they are difficult to see in observations, since Earth-based observation have very limited angular and temporal resolutions.

As mentioned before, the convection reaches two distinct pressure levels: one is close to the cloud base, and the other is near the 1 bar pressure level. The top cloud layer reaches a lower pressue in the region south of the jet, compared to the north. At this jet peak, there is a large wind shear near the 800 hPa level, and the large gradient in zonal wind speeds with latitude. Due to thermal wind balance, this produces a large meridional gradient where the south is cool and the north is warm. The cooler air therefore allows the convection to reach a lower pressure. This results in an interesting aspect, where convection near the southern flank of the jet can reach the ammonia cloud level, while the north does not. Indeed, observations show that the plumes are located on the southern flank of the jet \citep{SanchezLavega2017}, and these are reciprocated in our model. 


For the convecting storms, the water abundance is the primary factor in determining the strength of the storms. We see a clear increase in both convective mass and energy flux and storm frequency, as expected. With the water abundance at $0.5\times$, there are barely any convective cloud signatures. For the $1\times$ and $2\times$ cases, convection is predominantly tied to the jet and uses the wave over the jet as a guide for convection. The key difference is, however, in the distribution of storms, both zonally and meridionally. While the storms in the $1\times$ cases are more strongly located within the jet, storms in the $2\times$ cases are scattered nearly uniformly throughout the atmosphere. 

Furthermore, the $2\times$ cases reveal a nearly constant convective cloud below the jet, while the $1\times$ cases are much more sporadic and not as structured. A lack of a constant cloud cover below the jet is much more consistent with observations, which reveals regions of dark cloud clearing below the anticyclonic wave. Indeed, the advection of convective water storms form a constant layer near the 1-bar level in the $2\times$ cases in our model, which are not observed on Jupiter. Rather, the clearing between the ammonia arcs in observations, is, in fact, much more prominent than our models show for the $2\times$ cases. 

Note also, that the lack of convection here is also due to the trigger mechanism being different compared to most numerical convective adjustment schemes on Earth, who define a critical value of the cloud work function, or convective available potential energy. In Earth models, if the CAPE within that grid cell exceeds the critical value, then convection would work to bring it back. On Jupiter, this value is unknown, and therefore, we are likely to get more false negatives for convective triggers, since we always require the value of CAPE to constantly increase for convection to be triggered. The critical value of CAPE/CWF is an avenue that will need to be addressed, when better vertical profiles of the jovian atmosphere are available from observations of convection on Jupiter. This will provide more robust convective triggers in our model.

Studies of convective dynamics on Jupiter reveal a large range of convective energies. \citet{Hueso2001} find storms that have vertical velocities of up to $\sim150$ m/s for a $1\times$ solar water composition, but this sharply drops to about $30$ m/s for lower initial relative humidity. This corresponds to a CAPE of between 450-11000 J/kg. The larger end is similar to the peak vertical velocity obtained by \citet{Inurrigarro2020} in their modeling of the STB Ghost feature. This is likely an upper limit since they calculate the maximum vertical velocity based on the predicted temperature increase from latent heat release, as they do not directly model latent heat release. 

Our CAPE values of $\sim 500-1000$ J/kg and corresponding maximum vertical velocity values of $w \sim 30-50$ m/s falls in the lower end of this range, and is comparable to \citeauthor{Hueso2001}'s $50\%$ initial relative humidity case. It should be noted that we do not manually trigger convection in our model, as these studies do, but allow the model to evolve and trigger convection naturally through dynamical instabilities. It is possible that the instability in our modelled region is much weaker than that of \citet{Hueso2001} or \citet{Inurrigarro2020}, since they study specific convective plumes, which lead to large scale structures, requiring a the much higher convective energy. Furthermore, entrainment and the vertical water mixing ratio profiles complicate the vertical profiles for convection. In that regard, our results here likely predict the lower end of energy regime for water-based convection on Jupiter, since we fully treat the variations in vertical updraft profiles in our moist convective module for small storms that are driven only by dynamical instabilities. Large storms, which would likely cover the larger end of the energy regime, and are possibly triggered by heat pulses in the deep atmosphere are difficult to handle since they involve several unconstrained parameters (e.g., power, size, duration). A treatment of such storms will be part of a follow-up study with this convective module. 

While it is possible that the inclusion of the ammonium hydrosulfide layer would affect the dynamics of the upwelling water cloud in the 1-2 bar region, it is unlikely that the increased volatile mass alone would inhibit convection, since this requires a very high concentration of sulphur \citep{Nakajima2019}. Rather, it is more likely that water abundance in this region is less than $2\times$ solar. Indeed, this is also consistent with the energy flux for the different simulations, as the $2\times$ solar cases show persistent moist convective activity that releases about $80$ W/m$^2$ of energy near the jet region, which is roughly $10\times$ the average flux on Jupiter \citep{Li2018}. For the $1\times$ cases, the average flux peaks at roughly $10$ W/m$^2$, but these convective storms are shallow and do not constantly reach the ammonia cloud level.  \citet{Gierasch2000} found that convective storms generally carry an energy flux of 3.3 W/m$^2$ based on the frequency of lightning detected by {\it Galileo}. The convective energy flux from our tests are larger than this estimate by about twice for the $1\times$ case and about four times for the $2\times$ cases, which we attribute to the fact this is a region of large instability, which promotes more convective activity.

Therefore, given the lack of observations of a persistent water cloud near the 1~bar level, and the large discrepancy in the energy flux, we can infer that the water abundance on Jupiter, at this latitude, is likely closer to $1\times$ the solar [O/H] ratio. This value is consistent with observations near the Great Red Spot \citep{Bjoraker2018}, who obtained a value of at least $\sim1.1\times$ solar, and with the lower end of the results from the equatorial region \citep{Li2020}, who determined a value of $2.7^{+2.4}_{-1.7}$ solar. From our testing in Section~\ref{sec:2dsims}, we find that increasing the water abundance directly increases the frequency and intensity of convection until about $5\times$ the solar value. Therefore, from our 3D simulations, we can infer that an atmosphere with water abundance greater than $2\times$ solar would produce much more frequent and powerful convective storms that we have modeled here. In that case, from our analysis, it is unlikely that the water abundance in this region is much greater than $2\times$ the solar value. That said, we do not include mixed-phase microphysics \citep{Aglyamov2021,Guillot2020a}, and thus further tests are required in order to further constrain this upper limit.

However, constraining the water abundance in our model to a higher precision would require a better understanding of the deep volatile structure for both water and ammonia, and a description of the meridional gradient within the modeled region. {\it Juno}'s microwave radiometer (MWR) data shows that the jet region has ammonia enrichment near the 1-bar level \citep{Li2017}, but the water distribution is currently unknown, due to the MWR data being highly degenerate with volatile concentration and temperature. Regardless, there is likely a non-negligible meridional variations in the water abundance, and thus, further modeling studies of this region should be wary of the sensitivity of the initial volatile distribution to storm frequency profile.

\section*{Data Availability}
The data underlying this article is uploaded to Zenodo (DOI: \href{https://dx.doi.org/10.5281/zenodo.6210048}{10.5281/zenodo.6210048}). The data is split into three folders for each water abundance, and further into three \texttt{tar.gz} files, for each perturbation. Each tarball contains the model output and initialization, both in \texttt{netCDF4} format, for that model run. 

\section*{Acknowledgements}
We acknowledge support by the NASA's Early Career Fellowship (Grant No. 80NSSC18K0183), NASA's Solar System Workings (Grants No. NNX16AQ0), NASA's Cassini Data Analysis (Grant No. 80NSSC19K0198) and Future Investigators in NASA Earth and Space Science and Technology (Grant No. 80NSSC19K1541) programs. We also thank Dr. Timothy Dowling, the chief designer of EPIC, for his help in implementing the scheme, and Drs. Saida Caballero-Nieves, J{\'e}r{\'e}my Riousset and Steven Lazarus for their valuable inputs. We would also like to thank the two anonymous reviewers, whose comments have greatly improved the quality and clarity of the manuscript. 

\appendix

\section{Formalism of the RAS convective scheme}

\subsection{Buoyancy and vertical updraft profile}
With the given updraft parameterization,

\begin{equation}
	\label{eq:eta_RAS_2}
	\eta(\lambda, z) = 1 + \lambda \zeta + \lambda^2 \xi,
\end{equation}
and the definitions of the static energies, it is simple to define the vertical energy profile for a moist parcel within the cloud, as,

\begin{equation}
    \label{eq:h_profile}
    \dfrac{\partial h}{\partial z} = \dfrac{d\eta}{dz} \henv,
\end{equation}
where the right hand side denotes the change in the parcel's moist static energy from entraining dry air from the surrounding atmosphere. 

Similarly, it is possible to determine the total cloud mass profile,

\begin{equation}
    \dfrac{\partial \qt}{\partial z} = \dfrac{d\eta}{dz} \qtb,
\end{equation}
where $\qt = \sum_p q_p$ is the total moisture mass, including all phases $p$ (five in our model - vapor, cloud ice, cloud liquid, snow and rain). We can separate the cloud phases from the vapor by substituting,

\begin{equation}
	\label{eq:qc_profile}
	\dfrac{d(\eta \qt)}{dz} = \dfrac{d(\eta q)}{dz} + \dfrac{d(\eta \qc)}{dz} \Rightarrow 
	\dfrac{d(\eta \qc)}{dz} = \dfrac{d\eta}{dz} \qtb - \dfrac{d(\eta q)}{dz},
\end{equation}
where $q$ is strictly the vapor specific humidity, and $\qc$ is the total cloud condensate specific humidity (summed over all non-vapor phases). To determine the updraft vapor profile $q(z)$, we assume that the cloud is saturated. Therefore, we can approximate, similarly to \citetalias{ArakawaSchubert1974},

\begin{align}
	\label{eq:q_upd_profile}
	q(z) = q^*( T(z), p) \approx & \overline{q}^* + \left(\dfrac{dq^*}{dT}\right)\left( T - \overline{T}\right) \nonumber \\
	\approx & \overline{q}^* + \dfrac{1}{\cp} \left(\dfrac{dq^*}{dT}\right) (s - \overline{s}),
\end{align}
since $s - \overline{s} = \cp(T - \overline{T})$, and $T(z)$ is the updraft temperature profile. From Eq.~\eqref{eq:hstar}, with $h$ denoting the updraft value,

\begin{equation}
	h - \hsat = \cp ( T - \overline{T}) + L(q - \overline{q}^*) = s - \overline{s} + \dfrac{L}{\cp} \left(\dfrac{dq^*}{dT}\right) (s - \overline{s}).
\end{equation}
Therefore,

\begin{equation}
	\label{eq:s_diff_profile}
	s - \overline{s} = \dfrac{1}{1+\gamma} (h - \hsat),
\end{equation}
and, thus,

\begin{equation}
	\label{eq:q_diff_profile}
	q - \qsat = \dfrac{\gamma}{1 + \gamma}\dfrac{1}{L} (h - \hsat),
\end{equation}
where,

\begin{equation}
	\label{eq:gamma_sat}
	\gamma = \dfrac{L}{\cp} \left(\dfrac{dq^*}{dT}\right)
\end{equation}
With the vertical profiles of the moisture and static energies determined, we can determine the buoyancy profile in Eq~\ref{eq:buoyancy_main}, where the virtual temperature can be written in terms of the specific humidity, as,

\begin{equation}
	T_{\rm v} = T(1 + \nu q),
\end{equation}
where,

\begin{equation}
	\nu_i = \dfrac{R_i}{\Rd} - 1 + \sum_{j,j\neq i} \dfrac{w_j}{w_i}\left(\dfrac{R_j}{\Rd} - 1\right)
\end{equation}
for the $i$th species, with $R_i$ and $R_j$ being the specific gas constant for species $i$ and $j$,
respectively, and $\Rd$ being the specific gas constant for dry air. Using this, we can define the virtual 
dry static energy, $s_v$, using,

\begin{equation}
	s_v = \cp T_{\rm v} + \Phi = s + \nu \cp T q.
\end{equation}
Therefore, we can write the difference in virtual temperature as, 

\begin{align}
	\cp (T_{\rm v} - \Tvb) = & s_v - \overline{s_v} \nonumber \\
	= & s + \nu \cp T q - \overline{s} - \nu \cp \overline{T} \overline{q} \approx s \nonumber \\
	& \quad{}- \overline{s} + \nu \cp \overline{T} ( q - \overline{q}).
\end{align}
Using Eqs~\eqref{eq:s_diff_profile}, \eqref{eq:q_diff_profile} and \eqref{eq:gamma_sat}, we get,

\begin{equation}
	\cp (T_{\rm v} - \Tvb) =  \dfrac{1}{\tilde{L}} \left(h - \hstst \right)
\end{equation}
where,

\begin{equation}
	\label{eq:h_top_boundary}
	\hstst = \hsat - \dfrac{\nu \tilde{L}}{1 + \nu \qsat}(\qsat - \overline{q}),
\end{equation}
defines a virtual moist saturated static energy for the environment, and

\begin{equation}
	\tilde{L} = \dfrac{(1+\gamma)\cp \Tvb}{1 + \gamma \nu \cp \overline{T}/L}.
\end{equation}
Note also that $h=\hstst$ satisfies the top boundary condition when mass loading effects are ignored.
Physically, this means that the moist adiabatic profile of the updraft must match the virtual environmental 
saturated value at the cloud top to maintain neutral buoyancy. 
Consequently, the buoyancy can be written as,

\begin{equation}
	\label{eq:buoyancy}
	B = \dfrac{g}{\tilde{L}}\left[ h - \hstst - \tilde{L}( \qc - \qcb) \right].
\end{equation}
Since the cloud top must be neutrally buoyant, we take $B(\zd) = 0$, giving, the cloud top
boundary condition,

\begin{equation}
	\label{eq:cloud_top_buoyancy_simplified}
	h(\zd) - \hstst(\zd) - \tilde{L} \left( \qc(\zd) - \qcb(\zd)\right) = 0
\end{equation}

\subsection{Cloud top buoyancy condition in discrete form}

To determine $\lambda$, we need to solve for the quadratic profiles $\qc$ and $h$. 
We can write, from Eq.~\eqref{eq:qc_profile}, over a short interval $d z$,

\begin{equation}
	\label{eq:detaqc}
	d(\eta \qc) = \qt d\eta - d(\eta q),
\end{equation}
and, from Eq.~\eqref{eq:h_profile}

\begin{equation}
	d(\eta h) = \henv d(\eta),
\end{equation}
and, from Eq.~\eqref{eq:q_upd_profile},

\begin{equation}
	\eta q = \eta \qsat + \dfrac{\gamma}{L(1+ \gamma)} \eta (h - \hsat).
\end{equation}
In a discrete model, taking the finite difference, we get, from Eq.~\eqref{eq:detaqc},

\begin{align}
	\eta_{k-\frac{1}{2}} \quc_{k-\frac{1}{2}} - \eta_{k+\frac{1}{2}} \quc_{k+\frac{1}{2}} = &
	\qut_k (\eta_{k-\frac{1}{2}} - \eta_{k+\frac{1}{2}}) \nonumber \\
	& \quad{} - (\eta_{k-\frac{1}{2}} q_{k-\frac{1}{2}} - \eta_{k+\frac{1}{2}} q_{k+\frac{1}{2}}),
\end{align}
moving the cloud $C$ and total $T$ subscripts to superscripts for readability, where $k$ denotes the $k$-th vertical layer. Since $\quc$ depends on $q$, we can similarly write $q$ discretely as,

\begin{equation}
	\label{eq:q_upd_discrete_step1}
	\eta_{k-\half} q_{k-\half} = \eta_{k-\half} \qsat_{k-\half} + 
	\dfrac{\gamma_{k-\half}}{L(1+ \gamma_{k-\half})}\eta_{k-\half}\left( h_{k-\half} - \hsat_{k-\half}\right).
\end{equation}
Plugging in the discrete relation for $h$,

\begin{equation}
	\eta_{k-\half} h_{k-\half} = \eta_{k+\half} h_{k+\half} + \henv_{k} \left(\eta_{k-\half} - \eta_{k+\half}\right),
\end{equation}
we get,

\begin{align}
	\label{eq:h_upd_discrete}
	\eta_{k-\half} h_{k-\half} 
     = & \henv_K + \sum_{j=K-1}^{k} (\eta_{j-\half} - \eta_{j+\half}) \henv_j, \\						   
	 = &\henv_K + \sum_{j=K-1}^{k} \left[\lambda (\zeta_{j-\half} - \zeta_{j+\half}) \nonumber \right.\\
	 & \left. \quad{}+ \lambda^2 (\xi_{j-\half}  - \xi_{j+\half})\right] \henv_j \label{eq:updraft_h_value}
\end{align}
where we assume that the base value of $h$ is equal to the atmospheric value at the cloud base. 
Plugging this into Eq.~\eqref{eq:q_upd_discrete_step1}, we get

\begin{equation}
	\label{eq:q_upd_discrete}
	\eta_{k-\half} q_{k-\half} = C_{k-\half} + D_{k-\half} \lambda + E_{k-\half} \lambda^2,
\end{equation}
where $C$, $D$ and $E$ depend only on environment parameters. They are given by,

\begin{subequations}
	\begin{align}
	C_{k-\half} = & \: \qsat_{k-\half} -\dfrac{\gamma_{k-\half}}{L\left(1 + \gamma_{k-\half}\right)}\left(\hsat_{k-\half} - \henv_{K}\right)\label{eq:C_upd},\\
	D_{k-\half} = & \: \zeta_{k-\half} \left[ \qsat_{k-\half} - \dfrac{\gamma_{k-\half}}{L\left(1 + \gamma_{k-\half}\right)} \hsat_{k-\half}\right] \nonumber\\
				  & \: + \dfrac{\gamma_{k-\half}}{L\left(1+\gamma_{k-\half}\right)} \sum_{j=K-1}^k \left(\zeta_{j-\half} - \zeta_{j+\half}\right) \henv_{j}, \label{eq:D_upd}\\
	E_{k-\half} = & \: \xi_{k-\half} \left[ \qsat_{k-\half} - \dfrac{\gamma_{k-\half}}{L\left(1 + \gamma_{k-\half}\right)} \hsat_{k-\half}\right] \nonumber\\
				  & \: + \dfrac{\gamma_{k-\half}}{L\left(1+\gamma_{k-\half}\right)} \sum_{j=K-1}^k \left(\xi_{j-\half} - \xi_{j+\half}\right) \henv_{j} \label{eq:E_upd}.
	\end{align}
\end{subequations}

Plugging $q$ into the discrete equation for $\quc$, we get,
\begin{equation}
	\label{eq:qC_upd_discrete}
	\eta_{k-\half} \quc_{k-\half} = \eta_{k+\half} \quc_{k+\half} + F_k + G_k \lambda + H_k \lambda^2,
\end{equation}
where $F$, $G$ and $H$ are quantities that depend on environment parameters, and are given by,

\begin{subequations}
	\begin{align}
	F_k  = & \: C_{k+\half} - C_{k-\half} \\
	G_k  = & \: D_{k+\half} - D_{k-\half} + \qut_k \left(\zeta_{k-\half} - \zeta_{k+\half}\right)\\
	H_k  = & \: E_{k+\half} - E_{k-\half} + \qut_k \left(\xi_{k-\half} - \xi_{k+\half}\right)
\end{align}
\end{subequations}
Therefore, we can write the cloud updraft value at the top as,
\begin{equation}
	\eta_i \quc_i = \eta_{i+\half} \quc_{i+\half} + F_i + G_i \lambda + H_i \lambda^2,
\end{equation}
where we account only for the half-layer at the cloud top,

\begin{subequations}
	\begin{align}
	F_i  = & \: C_{i+\half} - C_{i}, \\
	G_i  = & \: D_{i+\half} - D_{i} + \qut_i \left(\zeta_{i} - \zeta_{i+\half}\right),\\
	H_i  = & \: E_{i+\half} - E_{i} + \qut_i \left(\xi_{i} - \xi_{i+\half}\right).
\end{align}
\end{subequations}
Adding these equations together, we can write,

\begin{equation}
	\label{eq:updraft_qc_top_value}
	\eta_i \quc_i = \sum_{j=K-1}^i \left(F_j +  G_j\lambda + H_j \lambda^2 \right) 
	= \tilde{F} + \tilde{G} \lambda + \tilde{H} \lambda^2,
\end{equation}

Multipyling the buoyancy condition (Eq.~\eqref{eq:cloud_top_buoyancy}) by $\eta_i$ and combining the 
the updraft $h$ profile (Eq.~\eqref{eq:updraft_h_value}) and Eq.~\eqref{eq:updraft_qc_top_value}, we get,

\begin{equation}
	\label{eq:RAS_discrete_lambda_eq}
	a \lambda^2 + b\lambda + c = 0,
\end{equation}
where $a$, $b$ and $c$ are functions of the environmental variables, and are given by,

\begin{subequations}
	\begin{align}
	a  = \: & \sum_{j=K-1}^{i+1} \left(\xi_{j-\half} -             \xi_{j+\half}\right) \henv_j \nonumber \\
		& + \left( \xi_i - \xi_{i+\half}\right) \henv_i - \xi_i \hstst_i \nonumber\\ 
		& - \tilde{L}_i (\tilde{H} - \xi_i q_i)\\ 
	b  = & \: \sum_{j=K-1}^{i+1} \left(\zeta_{j-\half} - \zeta_{j+\half}\right) \henv_j \nonumber \\
		& + \left( \zeta_i - \zeta_{i+\half}\right) \henv_i - \zeta_i \hstst_i \nonumber \\
		& - \tilde{L}_i (\tilde{G} - \zeta_i q_i)\\ 
	c  = & \: \henv_K - \hstst_i - \tilde{L}_i (\tilde{F} - q_i),
	\end{align}
\end{subequations}

\subsection{Large scale tendencies}
Using the updraft mass flux profile defined above, we can now calculate the change in environment (i.e., temperature and moisture content) as a result of the upwelling. 
Let $\sigma_i$ be the sub-grid, fractional area convered by each convecting cloud $i$, within a grid cell. Therefore, within a model layer, 

\begin{equation}
	\senv = \sum_i \sigma_i s_i + (1 - \sigma) \tilde{s},
\end{equation}
where $\tilde{s}$ defines the intra-cloud value of $\senv$ (i.e. the cloud free regions in between the updrafts) and $\sigma = \sum_i \sigma_i$. 
The evolution of $\tilde{s}$ is from 
\begin{itemize}
	\item large scale horizontal advection, i.e., $\rho \mathbf{v} \cdot \nabla s $
	\item vertical advection of $\tilde{s}$, i.e., $\rho w \dfrac{\partial s}{\partial z}$ 
\end{itemize}
Between the convecting clouds, the vertical mass flux $\tilde{M}$
is the difference between the large scale motion $\rho \overline{w}$ and cumulus updraft $M_u = \sum_i M_{b,i}$.
Since this generally corresponds to downdrafts, we define $-\tilde{M} = M_u - \rho \overline{w}$. 
Therefore, the energy conservation relation can be written as,
\begin{equation}
	(1 - \sigma)\rho \dfrac{\partial \senv(z)}{\partial t} 
	= -\overline{\rho \mathbf{v}} \cdot \nabla \tilde{s} - \tilde{M} \dfrac{\partial \tilde{s}}{\partial z}.
\end{equation}
Assuming that $\sigma \ll 1$, which is generally true for jovian convective clouds, given our model grid size of ${\sim} 100$ km,

\begin{subequations}
\begin{align}
	(1 - \sigma)\rho \dfrac{\partial \senv(z)}{\partial t} 
	 = & \:- \tilde{M} \dfrac{\partial \tilde{s}}{\partial z} - \overline{\rho \mathbf{v}} \cdot \nabla \tilde{s},\\[0.5em]
	 = & \: \tilde{M_u} \dfrac{\partial \tilde{s}}{\partial z} - \overline{\rho w} \dfrac{\partial \tilde{s}}{\partial z} 
	 - \overline{\rho \mathbf{v}} \cdot \nabla \tilde{s},\\[0.5em]
	 = & \: \tilde{M_u}\dfrac{\partial \tilde{s}}{\partial z} - \overline{\dfrac{\partial \tilde{s}}{\partial t}}\\[0.5em]
	 = & \: \dfrac{\partial (M_u \tilde{s})}{\partial z} - \tilde{s}\dfrac{\partial M_u}{\partial z} - \overline{\dfrac{\partial \tilde{s}}{\partial t}}
\end{align}
\end{subequations}
In the limit of $\sigma\rightarrow0$, we get, $\tilde{s} = \senv$, and therefore,

\begin{equation}
	\label{eq:s_tendency}
	\rho \dfrac{\partial \senv}{\partial t} = \dfrac{\partial (M_u \senv)}{\partial z} - \senv \dfrac{\partial M_u}{\partial z} + F_{\rm LS},
\end{equation}
where the first term represents vertical advection, the second represents entrainment and detrainment by clouds,
and the last is large scale advection. With the hydrostatic approximation, $\rho^{-1} dp = - g dz$, we get,

\begin{equation}
	\label{eq:s_tendency_pressure}
	\dfrac{\partial \senv}{\partial t} = -g \Mb \left[\dfrac{\partial (\eta\senv)}{\partial p} - \senv \dfrac{\partial \eta}{\partial p}\right] = \Mb \Gamma_s(z).
\end{equation}
Similarly, we write,

\begin{equation}
	\label{eq:h_tendency_pressure}
	\dfrac{\partial \henv}{\partial t} = -g \Mb \left[\dfrac{\partial (\eta\henv)}{\partial p} - \henv \dfrac{\partial \eta}{\partial p}\right] = \Mb \Gamma_h(z).
\end{equation}
Here, $\Gamma_s$ and $\Gamma_h$ define the rate of change of static energy per unit mass flux, for each cloud type. 
The net effect on the atmosphere is the sum of effects of all clouds. Here, $dh/dt$ defines the net heating rate on the atmosphere from both the convective upwelling and the latent heat release. 

Since the vertical flux profile is derived from mass conservation and energy conservation, we can simply use $\Gamma_s$ and $\Gamma_h$, and Eqs~\eqref{eq:s} and~\eqref{eq:h} to derive the grid scale moisture tendency,

\begin{equation}
	\overline{q} = \dfrac{1}{L} (\henv - \senv).
\end{equation}
Therefore,

\begin{equation}
	\label{eq:cumulus_moisture_tendency}
	\dfrac{\partial \overline{q}}{\partial t} = \dfrac{\Mb}{L} \left(\Gamma_h - \Gamma_s\right).
\end{equation}

In the case of advection of other phases and species, we can rewrite, from Eq.~\eqref{eq:s_tendency_pressure},
\begin{equation}
	\dfrac{\partial \overline{q}_p}{\partial t} = -g \Mb \left[\dfrac{\partial (\eta\overline{q}_p)}{\partial p} - \overline{q}_p \dfrac{\partial \eta}{\partial p}\right] = \Mb \Gamma_p(z),
\end{equation}
where $p$ corresponds to phases and species other than the vapor involved in cumulus updrafts.  In the case of jovian simulations with EPIC, $p$ refers to ice and liquid water clouds, and ammonia vapor. Currently, we do not track cloud phases of other species, since there are no multi-phase effects included in EPIC's microphysics scheme. Therefore,
advecting cloud values of other species would simply result in additional computational cost, with no difference
in the final result. This would be more important, for example, after the implementation of NH$_4$SH in the model,
where the water convection would advect the NH$_4$SH clouds up to the ammonia layer.

\subsection{Energy conservation}
We determine the CAPE for a parcel of air that starts at the cloud base, and reaches
neutral stability at the cloud top by the integral of the buoyancy profile given in Eq.~\eqref{eq:buoyancy},

\begin{align}
	{\rm CAPE} = & \int_{z_{\rm base}}^{z_{\rm top}} \dfrac{g}{\tilde{L}}\left[ h - \hstst - \tilde{L}(\quc - \qucb )\right] dz \nonumber \\
	= & \int_{\Phi_{\rm base}}^{\Phi_{\rm top}} \dfrac{1}{\tilde{L}}\left[ h - \hstst - \tilde{L}(\quc - \qucb )\right] d\Phi
\end{align}
The CAPE defines the total potential energy in the atmosphere, and therefore, must be a conserved, quantity. Therefore, to conserve energy within the grid cell,

\begin{align}
	\left(\dfrac{d\CAPE}{dt}\right) = \left(\dfrac{d\CAPE}{dt}\right)_{\rm LS} + \left(\dfrac{d\CAPE}{dt}\right)_{\rm cu} 
	\approx 0 \nonumber \\
	\Rightarrow \left(\dfrac{d\CAPE}{dt}\right)_{\rm cu} \approx - \left(\dfrac{d\CAPE}{dt}\right)_{\rm LS},
\end{align}
where the subscripts LS and cu denote contribution from large scale motion and cumulus upwelling, respectively.

The change in CAPE from cumulus upwelling is given by the change in the buoyancy, $B$, due to the updraft. 
This is given by,

\begin{align}
	& \dfrac{dB}{dt} = -\dfrac{1}{\tilde{L}} \left(\dfrac{d\hstst}{dt} + \tilde{L} \dfrac{d\qucb}{dt}\right) -\dfrac{\Mb}{\tilde{L}} \left(\Gamma_{\hstst} + \tilde{L} \Gamma_c\right), \nonumber \\
	& \Rightarrow F = \dfrac{1}{\tilde{L}}\left(\Gamma_{\hstst} + \tilde{L} \Gamma_c\right)
\end{align}
where $\Gamma_{\hstst}$ is the normalized virtual moist static energy tendency (determined from $\Gamma_h$ and $\Gamma_s$) and $\Gamma_c$ is the cloud condensate tendency, both per unit mass flux.

\bibliographystyle{aasjournal}

\bibliography{ref}

\end{document}